\documentclass[1p]{elsarticle}
\usepackage{soul}
\usepackage{hyperref}
\usepackage{amsmath}
\usepackage{graphicx}
\usepackage{subfig}
\usepackage{array}
\usepackage{amsfonts}
\usepackage[dvipsnames]{xcolor}
\usepackage{lipsum}
\usepackage{color}
\usepackage{nomencl}
\usepackage{mathtools}
\usepackage{amssymb}
\usepackage{booktabs}

\journal{} 

\begin{document}
\begin{frontmatter}
\title{Finite Element Time-Domain Body-of-Revolution Maxwell Solver based on Discrete Exterior Calculus}
\author[ad1]{Dong-Yeop Na}
\ead{na.94@osu.edu}
\author[ad2]{Ben-Hur V. Borges}
\ead{benhur@sc.usp.br}
\author[ad1]{Fernando L. Teixeira}
\ead{teixeira.5@osu.edu}
\address[ad1]{ElectroScience Laboratory and department of Electrical and Computer Engineering, The Ohio State University, Columbus, OH 43212, USA}
\address[ad2]{Electrical and Computer Engineering Department, University of S\~{a}o Paulo, S\~{a}o Carlos, SP 13560-970, Brazil}

\begin{abstract}
{\color{black}
We present a finite-element time-domain (FETD) Maxwell solver for the analysis of body-of-revolution (BOR) geometries based on discrete exterior calculus (DEC) of differential forms and transformation optics (TO) concepts.
We explore TO principles to map the original 3-D BOR problem to a 2-D one in the meridian $\rho z$-plane
 based on a Cartesian coordinate system where the cylindrical metric   
 is fully embedded into the constitutive properties of an effective inhomogeneous and anisotropic medium that fills the domain.
The proposed solver uses a $(\text{TE}^{\phi}, \text{TM}^{\phi})$ field decomposition and an appropriate set of DEC-based basis functions on an irregular grid discretizing the meridian plane.
A symplectic time discretization based on a leap-frog scheme is applied to obtain the full-discrete marching-on-time algorithm.
We validate the algorithm by comparing the numerical results against analytical solutions for resonant fields in cylindrical cavities and against pseudo-analytical solutions for fields radiated by cylindrically symmetric antennas in layered media. 
We also illustrate the application of the algorithm for a particle-in-cell (PIC) simulation of beam-wave interactions inside a high-power backward-wave oscillator.
}
\end{abstract}
\end{frontmatter}

{\color{black}
\section{Introduction} \label{sec:introduction}
The solution of Maxwell's equations in  circularly symmetric or body-of-revolution (BOR) geometries is important for a plethora of applications involving analysis and design of microwave devices (e.g. cavity resonators, coaxial cables, waveguides, antennas, high-power amplifiers, etc.)~\cite{jin2015the, lee1993finite, bergmann97a, bergmann97b, wilkins1991numerical, greenwood1999finite, rui2010higher,Tierens2011,na2017axisymmetric}, electromagnetic scattering~\cite{Joseph1993,Putnam1984,greenwood1999novel,donnell13}, metamaterials~\cite{zhai2010finite}, and exploration geophysics~\cite{pardo2006simulation,novo2008comparison, novo2010three,hong2017novel, yang2017stable, fang2017through, hue2005three}, to name a few. 
Azimuthal field variations in BOR problems can be described by Fourier modal decomposition, with the modal field solutions reduced to a two-dimensional (2-D) problem in the meridian $\rho z$-plane.
Frequency-domain finite element (FE) Maxwell solvers for BOR problems have been developed in the past by discretizing the second-order vector wave equation using edge elements for either the electric or the magnetic field ~\cite{greenwood1999finite,rui2010higher,greenwood1999novel,zhai2010finite,wong1995axisymmetric} which avoids some of the pitfalls encountered when using scalar elements~\cite{Joseph1993}.

It is highly desirable to develop BOR FE solvers in the time domain as well.
Time-domain FE solvers are better suited for simulating broadband problems, for capturing transient processes such as those involved in beam-wave interactions
~\cite{teixeira2008time,moon2015exact,na2016local}, and for handling non-linear problems.
However, the use of the second-order vector wave equation as a starting point for a time-domain FE formulation, as done in frequency-domain Maxwell FE solvers, is inadequate. This is because the vector wave equation admits solutions of the form $t\nabla\phi$, which are not original solutions of Maxwell's equations and, even if not excited by (properly set) initial conditions, may 
emerge in the course of the simulation due to round-off errors and pollute the results for long integration times~\cite{chilton2007discrete}.
To avoid this problem, a mixed (basis) FE solver based directly on the first-order should be adopted in the time domain~\cite{he2005on,he2006sparse,donderici2008conformal,donderici2008mixed}.

In this paper, we present a mixed FE BOR solver for time-domain Maxwell's curl equations based on transformation optics (TO) \cite{teixeira1999differential,pendry2006controlling,he2007differential,borges2008an,ozgun2013software,ozgun2014cartesian}
and discretization principles based on the discrete exterior calculus (DEC) of differential forms~\cite{teixeira2008time,he2005on,Kettunen1998,teixeira1999lattice,arnold2006finite,Kettunen2007,kim2011parallel,teixeira2013differential,teixeira2014lattice,Chen2017JCP}.
We explore TO principles to map the original three-dimensional (3-D) BOR problem to an equivalent problem on the 2-D meridian plane where the resulting metric is not the cylindrical one but instead the Cartesian one (i.e., with no radial factors present). The cylindrical metric becomes fully embedded into the constitutive properties of an effective (artificial) inhomogeneous anisotropic medium that fills the entire domain. In this way, a Cartesian 2-D FE code can be retrofitted to this problem with no modifications necessary except to accommodate the presence of anisotropic media. 
Similar ideas have been explored in the past but restricted to the frequency-domain finite-difference (FD) context and to structured grids only~\cite{shyroki2007efficient}.  
In the FE context considered here, DEC principles are used to discretize Maxwell's equations on unstructured (irregular) grids using discrete differential (Whitney) forms~\cite{ he2007differential,Kettunen1998,Kettunen2007,he2007mixed,he2006geometric}. Unstructured grids permits a more flexible representation of irregular geometries and reduce the need for geometrical defeaturing.
In addition to the above advantages, the proposed formalism facilitates treatment of the coordinate singularity on the axis of symmetry ($z$ axis) because it does not require any modification of the basis functions for $\rho=0$ (otherwise necessary in prior BOR FE solvers~\cite{greenwood1999finite,greenwood1999novel,wong1995axisymmetric}). 
As detailed in the Appendix, the DEC formalism also facilitates implementation of perfectly matched layers (PML) to truncate the outer boundaries.
We validate the algorithm against analytical solutions for resonant fields in cylindrical cavities and against pseudo-analytical solutions for the radiated fields by cylindrically symmetric antennas in layered media.
We also illustrate the application of the algorithm to the simulation of wave-beam interactions in a high-power microwave backward-wave oscillator (BWO).}

{\color{black}
\section{Formulation}
\subsection{Exploration of transformation optics (TO) concepts}
Consider a  BOR object with symmetry axis along $z$, such as the waveguide structure depicted in Fig. \ref{fig:cyl_wg}.
\begin{figure}
    \centering	
    \includegraphics[width=3in]{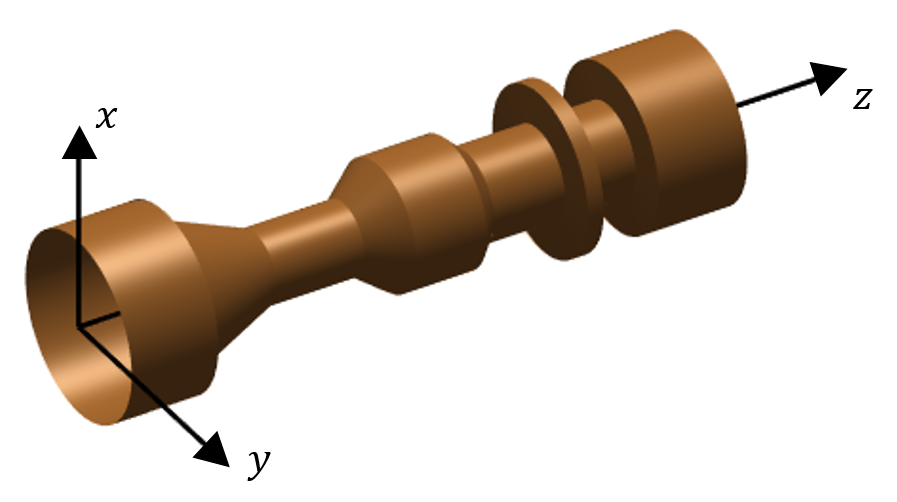}
\caption{Depiction of an axisymmetric structure.}
\label{fig:cyl_wg}
\end{figure}
It is well known that the vector operators (gradient, curl, and divergence) in cylindrical coordinates have additional metric scaling factors not present in Cartesian coordinates.
However, by exploiting TO concepts~\cite{teixeira1999differential,pendry2006controlling,teixeira1999lattice}, we can map the cylindrical-system Maxwell's curl equations to a Cartesian-like equations where the metric factors are embedded into {\it artificial} constitutive tensors.
{\color{black}For convenience we denote these calculations under the generic banner of TO but some of these ideas actually predate TO per se. They can be traced to earlier applications involving Maxwell's equations in BOR geometries and to Weitzenbock identities involving differential forms of different degrees~\cite{Kotiuga2} in cylindrical (polar) coordinates.}

Starting from Maxwell's equations in cylindrical coordinates, and considering artificial anisotropic permittivity and permeability tensors $\bar{\bar{\epsilon}}'$ and $\bar{\bar{\mu}}'$ of the form
\begin{eqnarray}
&\bar{\bar{\epsilon}}'=\bar{\bar{\epsilon}}\cdot \bar{\bar{\text{R}}}_{\epsilon}=\bar{\bar{\epsilon}}\cdot
\left[
\begin{matrix}
    \rho & 0 & 0 \\
    0 & \rho^{-1} & 0 \\
    0 & 0 & \rho
\end{matrix}
\right],
\\
&\bar{\bar{\mu}}'=\bar{\bar{\mu}}\cdot \bar{\bar{\text{R}}}_{\mu}=\bar{\bar{\mu}}\cdot
\left[
\begin{matrix}
    \rho^{-1} & 0 & 0 \\
    0 & \rho & 0 \\
    0 & 0 & \rho^{-1}
\end{matrix}
\right],
\end{eqnarray}
where the constitutive parameters of the original medium are given by
\begin{eqnarray}
~~~
\bar{\bar{\epsilon}}=
\left[
\begin{matrix}
    \epsilon_{\rho} & 0 & 0 \\
    0 &     \epsilon_{\phi} & 0 \\
    0 & 0 &     \epsilon_{z} 
\end{matrix}
\right],
~~
\bar{\bar{\mu}}=
\left[
\begin{matrix}
    \mu_{\rho} & 0 & 0 \\
    0 & \mu_{\phi} & 0 \\
    0 & 0 & \mu_{z} 
\end{matrix}
\right].
\nonumber
\end{eqnarray}
and using the following rescaling for the fields 
\begin{flalign}
&\mathbf{E}'=\bar{\bar{\text{R}}}_{\mathbf{E}}\cdot\mathbf{E}=
\left[
\begin{matrix}
    1 & 0 & 0 \\
    0 & \rho & 0 \\
    0 & 0 & 1
\end{matrix}
\right]
\cdot\mathbf{E},
\\
&\mathbf{D}'=\bar{\bar{\text{R}}}_{\mathbf{D}}\cdot\mathbf{D}=
\left[
\begin{matrix}
    \rho & 0 & 0 \\
    0 & 1 & 0 \\
    0 & 0 & \rho
\end{matrix}
\right]
\cdot\mathbf{D},
\\
&\mathbf{B}'=\bar{\bar{\text{R}}}_{\mathbf{B}}\cdot\mathbf{B}=
\left[
\begin{matrix}
    \rho & 0 & 0 \\
    0 & 1 & 0 \\
    0 & 0 & \rho
\end{matrix}
\right]
\cdot\mathbf{B},
\\
&\mathbf{H}'=\bar{\bar{\text{R}}}_{\mathbf{H}}\cdot\mathbf{H}=
\left[
\begin{matrix}
    1 & 0 & 0 \\
    0 & \rho & 0 \\
    0 & 0 & 1
\end{matrix}
\right]
\cdot\mathbf{H},
\end{flalign}
we can rewrite the resulting Maxwell's curl equations as 
\begin{flalign}
\nabla'\times\mathbf{E}'&=-\frac{\partial \mathbf{B}'}{\partial t},
\label{eq:FL}
\\
\nabla'\times\mathbf{H}'&=\frac{\partial \mathbf{D}'}{\partial t},
\label{eq:AL}
\\
\mathbf{D}'&=\bar{\bar{\epsilon}}'\cdot \mathbf{E}',
\\
\mathbf{B}'&=\bar{\bar{\mu}}'\cdot \mathbf{H}',
\end{flalign}
with
\begin{flalign}
\nabla'\times\mathbf{A}'=
\left|
\begin{matrix}
    \hat{\rho} & \hat{\phi} & \hat{z} \\
    \frac{\partial}{\partial \rho} & \frac{\partial}{\partial \phi} & \frac{\partial}{\partial z} \\
    A_{\rho}' & A_{\phi}' & A_{z}'
\end{matrix}
\right|.
\label{eq:A_curl_eqv}
\end{flalign}
The modified curl operator in the equivalent (primed) system seen in (\ref{eq:A_curl_eqv}) is devoid of any radial scaling and thus locally isomorphic to the Cartesian curl operator. 
}

{\color{black}
\clearpage
\subsection{Field decomposition}\label{subsec:decomposition}
We decompose the fields into two sets: $\text{TE}^{\phi}$- and $\text{TM}^{\phi}$-polarized fields, corresponding to $\left\{{E'}_{\rho}, {E'}_{z}, {B'}_{\phi}\right\}$ and $\left\{{E'}_{\phi}, {B'}_{\rho}, {B'}_{z}\right\}$, respectively.
In what follows, we use superscripts $^{\parallel}$ or $^{\perp}$ to denote fields transverse or normal to the 2-D meridian plane.
The $\text{TE}^{\phi}$ field components can be expressed as $\mathbf{E'}^{\parallel}$ and $\mathbf{B'}^{\perp}$ and the $\text{TM}^{\phi}$ as $\mathbf{E'}^{\perp}$ and $\mathbf{B'}^{\parallel}$.
In the DEC context, the electric field intensity, the magnetic flux density, the electric flux density, and the magnetic field intensity are likewise represented as 1-,  2-, 2-, and 1-forms\footnote{1- and 2-forms correspond to physical quantities naturally associated to line and surface integrals, respectively.} on the 3-D Euclidean space, respectively~\cite{teixeira1999lattice}.
For present analysis based on the meridian plane (a 2-D manifold), $\mathcal{E}^{\parallel}$ is transverse to the plane and still is represented as a 1-form. On the other hand, $\mathcal{E}^{\perp}$ should be represented as a 0-form since it is a point-based quantity on this manifold.
Likewise, although $\mathcal{B}^{\perp}$ is a 2-form in 3-D, $\mathcal{B}^{\parallel}$ is represented as a 1-form on the 2-D meridian plane (see Fig.~\ref{fig:pdm}).

\begin{figure}
    \centering
	\subfloat[\label{fig:pm}]{
    \includegraphics[width=1.8in]{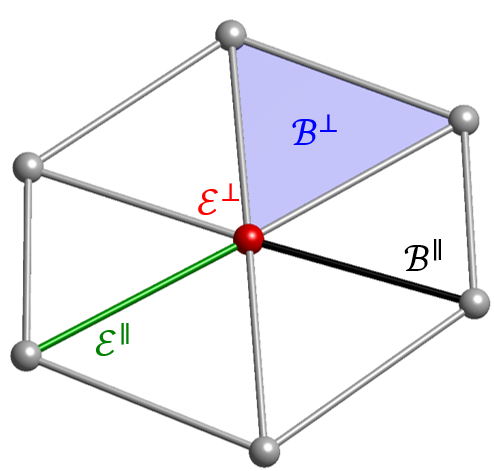}
    }
	\subfloat[\label{fig:dm}]{
    \includegraphics[width=2in]{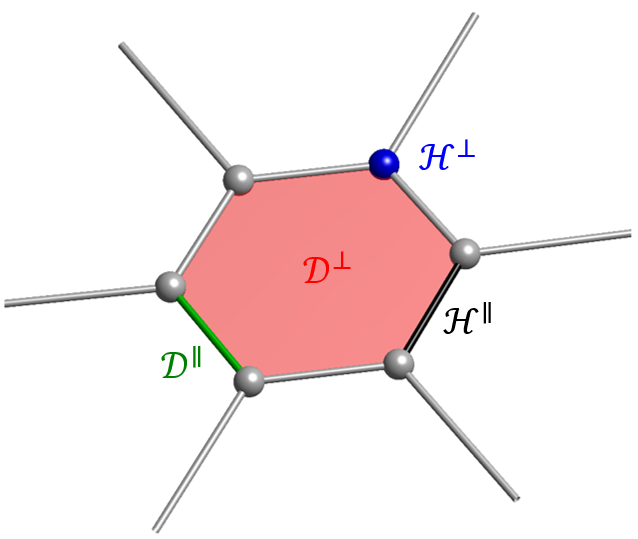}
    }
\caption{(2+1) setup for fields on (a) primal and (b) dual meshes at the meridian plane. The vertical axis is $\rho$ and the horizontal axis is $z$.}
\label{fig:pdm}
\end{figure}
}

{\color{black}
\clearpage
\subsection{Mixed FE time-domain BOR solver}
We factor the transverse (i.e. $\rho$ and $z$) and normal (i.e. $\phi$) variations of the polarization-decomposed Maxwell fields on the 2-D meridian plane as
\begin{flalign}
\mathbf{E'}\left(\rho,\phi,z,t\right)=\sum_{m=-M_\phi}^{M_\phi}\mathbf{E'}^{\parallel}_{m}\left(\rho,z,t\right)\Phi_{m}\left(\phi\right)+\sum_{m=-M_\phi}^{M_\phi}\mathbf{E'}^{\perp}_{m}\left(\rho,z,t\right)\Psi_{m}\left(\phi\right),
\label{eq:E_eqv_rep}
\\
\mathbf{B'}\left(\rho,\phi,z,t\right)=\sum_{m=-M_\phi}^{M_\phi}\mathbf{B'}^{\perp}_{m}\left(\rho,z,t\right)\Phi_{m}\left(\phi\right)+\sum_{m=-M_\phi}^{M_\phi}\mathbf{B'}^{\parallel}_{m}\left(\rho,z,t\right)\Psi_{m}\left(\phi\right),
\label{eq:B_eqv_rep}
\end{flalign}
where $M_{\phi}$ is the maximum order of the Fourier harmonics considered and 
\begin{flalign}
\Phi_{m}\left(\phi\right)=
\left\{
  \begin{array}{@{}ll@{}}
    \cos\left(m\phi\right), & \text{for}\ m<0 \\
    1, & \text{for}\ m=0 \\
    \sin\left(m\phi\right), & \text{for}\  m>0
  \end{array}\right.
,
\\
\Psi_{m}\left(\phi\right)=
\left\{
  \begin{array}{@{}ll@{}}
    \sin\left(m\phi\right), & \text{for}\ m<0 \\
    1, & \text{for}\ m=0 \\
    \cos\left(m\phi\right), & \text{for}\  m>0
  \end{array}\right.
.
\end{flalign}
Substituting (\ref{eq:E_eqv_rep}) and (\ref{eq:B_eqv_rep}) into (\ref{eq:FL}), by using the orthogonality between modes, i.e.
\begin{flalign}
\int_{0}^{2\pi} \Phi_{m}\left(\phi\right) \Phi_{n}\left(\phi\right) d\phi&=C_{m} \delta_{mn}, \\
\int_{0}^{2\pi} \Psi_{m}\left(\phi\right) \Psi_{n}\left(\phi\right) d\phi&=C_{m} \delta_{mn}, 
\end{flalign}
where  $C_{m}=\pi$ for $m\neq0$ and $C_{0}=2\pi$, we obtain the modal Faraday's law as
\begin{flalign}
{\nabla'}^{\parallel}\times \mathbf{E'}^{\parallel}_{m}\left(\rho,z,t\right)&=-\frac{\partial \mathbf{B'}^{\perp}_{m}\left(\rho,z,t\right)}{\partial t},
\label{eq:FL_eqv_TE_pol}
\\
{\nabla'}^{\parallel}\times \mathbf{E'}^{\perp}_{m}\left(\rho,z,t\right)&=-\frac{\partial \mathbf{B'}^{\parallel}_{m}\left(\rho,z,t\right)}{\partial t} 
{\color{black}
+ \left|m\right|\mathbf{E'}^{\parallel}_{m}\left(\rho,z,t\right)\times\hat{\phi},
}
\label{eq:FL_eqv_TM_pol}
\end{flalign}
for $m=-M_{\phi},...,M_{\phi}$, where ${\nabla'}^{\parallel}=\hat{\rho}{\partial}/{\partial \rho}+\hat{z}{\partial}/{\partial z}$.

We discretize (\ref{eq:FL_eqv_TE_pol}) and (\ref{eq:FL_eqv_TM_pol}) on the meridian plane using an unstructured mesh based on simplicial (triangular) cells and by expanding the fields in a mixed basis as scalar or vector proxies
of discrete differential forms (Whitney forms)~\cite{he2005on,teixeira1999lattice,teixeira2014lattice}.
In particular, the $\text{TE}^{\phi}$ field is expanded as
\begin{flalign}
\mathbf{E'}^{\parallel}_{m}\left(\rho,z,t\right)&=
\sum_{j=1}^{N_{1}}
\mathbb{E}_{j,m}^{\parallel}\left(t\right)\mathbf{W}_{j}^{(1)}\left(\rho,z\right),
\label{eq:E_bs_par}
\\
\mathbf{B'}^{\perp}_{m}\left(\rho,z,t\right)&=
\sum_{k=1}^{N_{2}}
\mathbb{B}_{k,m}^{\perp}\left(t\right) \mathbf{W}_{k}^{(2)}\left(\rho,z\right),
\label{eq:B_bs_perp}
\end{flalign}
where $\mathbf{W}_{q}^{(p)}$ is the
 vector proxy of a Whitney $p$-form $w_{q}^{(p)}$~\cite{moon2015exact} associated with the $q$-th $p$-cell 
 ($p=0,1,2$ for nodes, edges, and facets, respectively) on the grid, and $N_{p}$ is the total number of $p$-cells on the grid. The expressions for the Whitney forms and their proxies are provided in \ref{ap:Forms}.
Likewise, the $\text{TM}^{\phi}$ field is represented as
\begin{flalign}
\mathbf{E'}^{\perp}_{m}\left(\rho,z,t\right)&=
\sum_{i=1}^{N_{0}}\mathbb{E}_{i,m}^{\perp}\left(t\right) \hat{\phi} \, \,
{\text{W}}_{i}^{(0)}\left(\rho,z\right),
\label{eq:E_bs_perp}
\\
{\color{black}
\mathbf{B'}^{\parallel}_{m}\left(\rho,z,t\right)
}&
{\color{black}
=
\sum_{j=1}^{N_{1}}\mathbb{B}_{j,m}^{\parallel}\left(t\right) 
\mathbf{W}^{(\text{RWG})}_{j}\left(\rho,z\right)
}.
\label{eq:B_bs_par}
\end{flalign}
In what follows, we denote 
{\color{black}
$\mathbf{W}^{(1)}_{j}\times\hat{\phi}=\mathbf{W}_{j}^{(\text{RWG})}$}, since this expression recovers the so-called Rao-Wilton-Glisson (RWG) functions~\cite{RWG1982,warnickbook2008}~\footnote{In other words, 
$\mathbf{W}_{j}^{(\text{RWG})}$ is the Hodge dual of 
$\mathbf{W}^{(1)}_{j}$ in 2-D~\cite{Kettunen2007,teixeira2014lattice,Gillette2011}.}. 
Note that we use dummy index subscripts $i$, $j$, and $k$ to indicate the $i$-th node, $j$-th edge, and $k$-th face, respectively.
\begin{figure}[t]
     \centering	
	\subfloat[\label{fig:w_1}]{
     \includegraphics[width=1.75in]{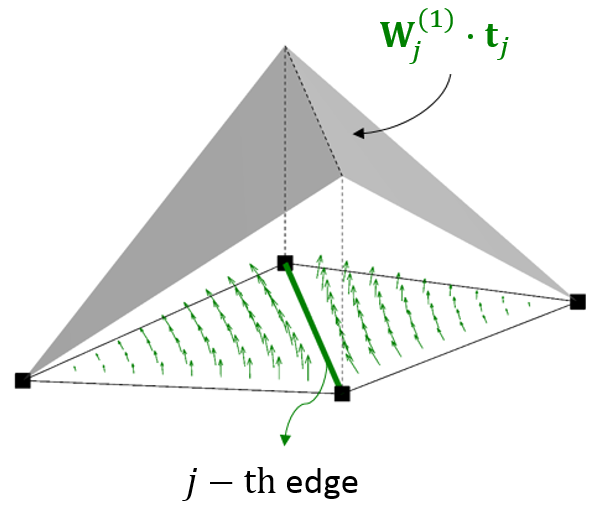}
	}
	\subfloat[\label{fig:w_2}]{
     \includegraphics[width=1.75in]{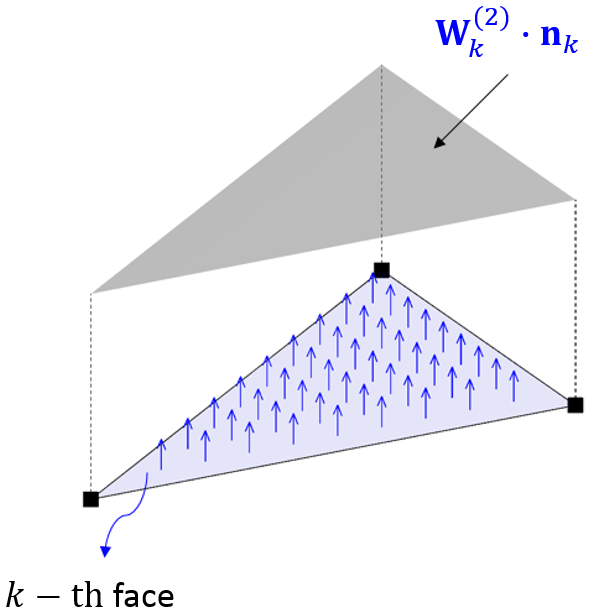}
	}
\\
	\subfloat[\label{fig:w_0}]{
	\includegraphics[width=1.75in]{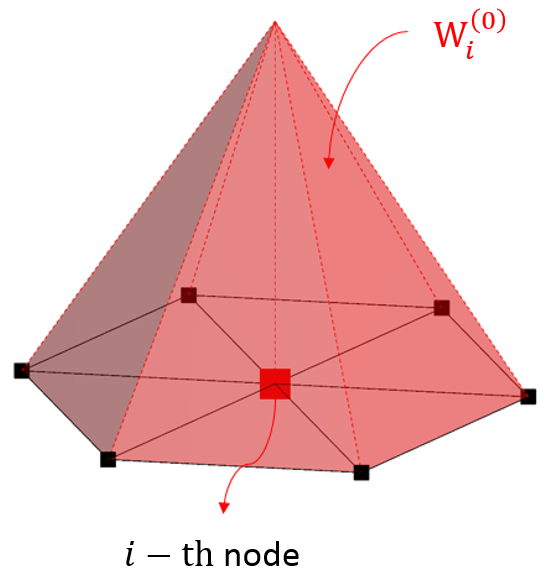}     
	}     
	\subfloat[\label{fig:w_RWG}]{
	\includegraphics[width=1.75in]{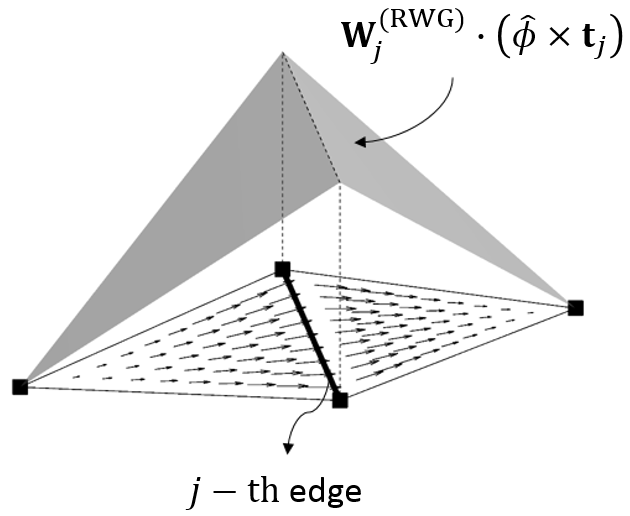}     
	}     
	\caption{Vector proxies of various degrees of Whitney forms on the mesh: (a) $\mathbf{W}^{(1)}_{j}$, (b) $\mathbf{W}^{(2)}_{k}$, (c) ${\text{W}}^{(0)}_{i}$, and (d) $\mathbf{W}^{(\text{RWG})}_{j}$. Note that $\mathbf{t}_{j}$ is a unit vector tangential to $j-$th edge and parallel to its direction and $\mathbf{n}_{k}$ is a unit vector normal to $k-$th face.}
\label{fig:bf}
\end{figure}
The various basis functions above are depicted in Fig. \ref{fig:bf}, see also~\cite{bossavit1988whitney,kotiuga2004electromagnetic}.

By substituting (\ref{eq:E_bs_par}) and (\ref{eq:B_bs_perp}) into (\ref{eq:FL_eqv_TE_pol}), and (\ref{eq:E_bs_perp}) and (\ref{eq:B_bs_par}) into (\ref{eq:FL_eqv_TM_pol}), we obtain the following equations
\begin{flalign}
&\sum_{j=1}^{N_{1}}
\mathbb{E}_{j,m}^{\parallel}\left(t\right) 
\left({\nabla'}^{\parallel}\times \mathbf{W}_{j}^{(1)}\right)
=-\frac{\partial}{\partial t}
\sum_{k=1}^{N_{2}}
\mathbb{B}_{k,m}^{\perp}\left(t\right) 
\mathbf{W}_{k}^{(2)}
\\
&
{\color{black}
\sum_{i=1}^{N_{0}}
\mathbb{E}_{i,m}^{\perp}\left(t\right) 
{\nabla'}^{\parallel}\text{W}_{i}^{(0)}
=-\frac{\partial}{\partial t}
\sum_{j=1}^{N_{1}}
\mathbb{B}_{j,m}^{\parallel}\left(t\right) \mathbf{W}_{j}^{(1)}
+\left|m\right|\sum_{j=1}^{N_{1}}
\mathbb{E}_{j,m}^{\parallel}\left(t\right)  \mathbf{W}_{j}^{(1)},
}
\end{flalign}
for $m=-M_{\phi},...,M_{\phi}$ and where we have used the fact that {\color{black}${\nabla'}^{\parallel} \times \left(  \hat{\phi} \text{W}_{i}^{(0)} \right) = \left({\nabla'}^{\parallel} \text{W}_{i}^{(0)}\right)\times\hat{\phi}$}.
The equations above can be recast using the exterior calculus of differential forms as
\begin{flalign}
\sum_{j=1}^{N_{1}}
\mathbb{E}_{j,m}^{\parallel}\left(t\right) 
\left({d'}^{\parallel}w_{j}^{(1)}\right)
&=-\frac{\partial}{\partial t}
\sum_{k=1}^{N_{2}}
\mathbb{B}_{k,m}^{\perp}\left(t\right) w_{k}^{(2)},
\label{eq:diff_FL_TE_pol}
\\
{\color{black}
\sum_{i=1}^{N_{0}}
\mathbb{E}_{i,m}^{\perp}\left(t\right) 
\left({d'}^{\parallel}w_{i}^{(0)}\right)
}
&
{\color{black}
=-\frac{\partial}{\partial t}
\sum_{j=1}^{N_{1}}
\mathbb{B}_{j,m}^{\parallel}\left(t\right) w_{j}^{(1)}
+\left|m\right|\sum_{j=1}^{N_{1}}
\mathbb{E}_{j,m}^{\parallel}\left(t\right) 
w_{j}^{(1)},
}
\label{eq:diff_FL_TM_pol}
\end{flalign}
where ${d'}^{\parallel}=d\rho \, {\partial}/{\partial \rho}+dz \, {\partial}/{\partial z}$ is the exterior derivative on the meridian plane.

Applying DEC principles, (\ref{eq:diff_FL_TE_pol}) can be paired to the 2-cells of the mesh and (\ref{eq:diff_FL_TM_pol}) to the 1-cells of the mesh  (see \ref{ap:DEC1})
so that, by invoking the generalized Stokes' theorem~\cite{he2005on,teixeira1999lattice,Kettunen2007,teixeira2014lattice,Chen2017JCP}  (see \ref{ap:DEC2}), the exterior derivative can be 
replaced by incidence operators on the mesh (see also \ref{ap:incidence}). Next, by discretizing the time derivatives using central-differences in a staggered manner  
(leap-frog time discretization)
we obtain the following update equations for Faraday's law
\begin{flalign}
\left[\mathbb{B}_{m}^{\perp}\right]^{n+\frac{1}{2}}&=
\left[\mathbb{B}_{m}^{\perp}\right]^{n-\frac{1}{2}}-\Delta t\left[\mathcal{D}_{\text{curl}}\right]\cdot\left[\mathbb{E}_{m}^{\parallel}\right]^{n},
\\
{\color{black}
\left[\mathbb{B}_{m}^{\parallel}\right]^{n+\frac{1}{2}}
}
&
{\color{black}
=\left[\mathbb{B}_{m}^{\parallel}\right]^{n-\frac{1}{2}}
-\Delta t \left(\left[\mathcal{D}_{\text{grad}}\right] \cdot\left[\mathbb{E}_{m}^{\perp}\right]^{n}
-\left|m\right| \left[\mathbb{E}_{m}^{\parallel}\right]^{n}\right),
}
\end{flalign}
where $\Delta t$ is a time step increment and the superscript $n$ indicates the time-step index. $\left[\mathcal{D}_{\text{curl}}\right]$ and $\left[\mathcal{D}_{\text{grad}}\right]$ are $N_{2} \times N_{1}$ and $N_{1} \times N_{0}$  incidence matrices, respectively, that encode the curl and the gradient operators on the FE mesh with elements in the set $\left\{-1,0,1\right\}$ (see \ref{ap:incidence}). The field unknowns are represented by the column vectors
  $\left[\mathbb{B}_{m}^{\perp}\right]=\left[\mathbb{B}_{m,1}^{\perp},...,\mathbb{B}_{m,N_{2}}^{\perp}\right]^T$, $\left[\mathbb{E}_{m}^{\parallel}\right]=\left[\mathbb{E}_{m,1}^{\parallel},...,\mathbb{E}_{m,N_{1}}^{\parallel}\right]^T$, $\left[\mathbb{B}_{m}^{\parallel}\right]=\left[\mathbb{B}_{m,1}^{\parallel},...,\mathbb{B}_{m,N_{1}}^{\parallel}\right]^T$, and $\left[\mathbb{E}_{m}^{\perp}\right]=\left[\mathbb{E}_{m,1}^{\perp},...,\mathbb{E}_{m,N_{0}}^{\perp}\right]^T$.

We proceed along similar lines for Ampere's law by expressing the $\mathbf{D'}$ and $\mathbf{H'}$ fields as
\begin{flalign}
\mathbf{D'}\left(\rho,\phi,z,t\right)=\sum_{m=0}^{M_\phi}\mathbf{D'}^{\parallel}_{m}\left(\rho,z,t\right)\Phi_{m}\left(\phi\right)+\sum_{m=0}^{M_\phi}\mathbf{D'}^{\perp}_{m}\left(\rho,z,t\right)\Psi_{m}\left(\phi\right),
\label{eq:D_eqv_rep}
\\
\mathbf{H'}\left(\rho,\phi,z,t\right)=\sum_{m=0}^{M_\phi}\mathbf{H'}^{\perp}_{m}\left(\rho,z,t\right)\Phi_{m}\left(\phi\right)+\sum_{m=0}^{M_\phi}\mathbf{H'}^{\parallel}_{m}\left(\rho,z,t\right)\Psi_{m}\left(\phi\right).
\label{eq:H_eqv_rep}
\end{flalign}
After substituting (\ref{eq:D_eqv_rep}) and (\ref{eq:H_eqv_rep}) to (\ref{eq:AL}), applying trigonometric orthogonality to the resulting equations, and matching the field components, we arrive at
\begin{flalign}
{\nabla'}^{\parallel}\times \mathbf{H'}^{\parallel}_{m}\left(\rho,z,t\right)&=\frac{\partial \mathbf{D'}^{\perp}_{m}\left(\rho,z,t\right)}{\partial t},
\label{eq:AL_TM_pol_vec}
\\
{\color{black}
{\nabla'}^{\parallel}\times \mathbf{H'}^{\perp}_{m}\left(\rho,z,t\right)
}
&
{\color{black}
=\frac{\partial \mathbf{D'}^{\parallel}_{m}\left(\rho,z,t\right)}{\partial t} - \left|m\right|\mathbf{H'}^{\parallel}_{m}\left(\rho,z,t\right)\times\hat{\phi}.
}
\label{eq:AL_TE_pol_vec}
\end{flalign}
As before, we discretize (\ref{eq:AL_TM_pol_vec}) and (\ref{eq:AL_TE_pol_vec}) on the 2-D meridian plane, the important difference being that the discretization for 
$\mathbf{D'}$ and $\mathbf{H'}$ is on the {\it dual} mesh
~\cite{he2007differential,teixeira1999lattice,teixeira2014lattice,Gillette2011}, as opposed to the FE (primal) mesh as done for $\mathbf{E'}$ and $\mathbf{B'}$.
In this way, we obtain
\begin{flalign}
\mathbf{D'}^{\parallel}_{m}\left(\rho,z,t\right)&=
\sum_{j=1}^{\tilde{N}_{1}}
\mathbb{D}_{j,m}^{\parallel}\left(t\right)\tilde{\mathbf{W}}_{j}^{(\text{RWG})}\left(\rho,z\right),
\label{eq:D_bs_par}
\\
\mathbf{H'}^{\perp}_{m}\left(\rho,z,t\right)&=
\sum_{i=1}^{\tilde{N}_{0}}
\mathbb{H}_{i,m}^{\perp}\left(t\right) \hat{\phi} \tilde{\text{W}}_{i}^{(0)}\left(\rho,z\right),
\label{eq:H_bs_perp}
\\
\mathbf{D'}^{\perp}_{m}\left(\rho,z,t\right)&=
\sum_{k=1}^{\tilde{N}_{2}}\mathbb{D}_{k,m}^{\perp}\left(t\right) \tilde{\mathbf{W}}_{k}^{(2)}\left(\rho,z\right),
\label{eq:D_bs_perp}
\\
\mathbf{H'}^{\parallel}_{m}\left(\rho,z,t\right)&=
\sum_{j=1}^{\tilde{N}_{1}}\mathbb{H}_{j,m}^{\parallel}\left(t\right) \tilde{\mathbf{W}}_{j}^{(1)}\left(\rho,z\right).
\label{eq:H_bs_par}
\end{flalign}
where we use the tilde $\tilde{}$ to denote quantities associated with the dual mesh.
Similar to the discrete counterparts of Faraday's law, by substituting (\ref{eq:D_bs_par}) and (\ref{eq:H_bs_perp}) into (\ref{eq:AL_TM_pol_vec}) and (\ref{eq:D_bs_perp}) and (\ref{eq:H_bs_par}) into (\ref{eq:AL_TE_pol_vec}) and by applying DEC principles and a leap-frog time discretization to the resulting equations, we obtain the discrete representations of Ampere's law as
\begin{flalign}
\left[\mathbb{D}_{m}^{\perp}\right]^{n+1}&=
\left[\mathbb{D}_{m}^{\perp}\right]^{n}+\Delta t\left[\tilde{\mathcal{D}}_{\text{curl}}\right]\cdot\left[\mathbb{H}_{m}^{\parallel}\right]^{n+\frac{1}{2}},
\\
{\color{black}
\left[\mathbb{D}_{m}^{\parallel}\right]^{n+1}
}
&
{\color{black}
=\left[\mathbb{D}_{m}^{\parallel}\right]^{n}
+\Delta t \left(\left[\tilde{\mathcal{D}}_{\text{grad}}\right] \cdot \left[\mathbb{H}_{m}^{\perp}\right]^{n+\frac{1}{2}}-\left|m\right| \left[\mathbb{H}_{m}^{\parallel}\right]^{n+\frac{1}{2}}\right),
}
\end{flalign}
where $\left[\tilde{\mathcal{D}}_{\text{curl}}\right]$ and $\left[\tilde{\mathcal{D}}_{\text{grad}}\right]$ are incidence matrices on the dual mesh, with sizes $\tilde{N}_{2} \times \tilde{N}_{1}$ and $\tilde{N}_{1} \times \tilde{N}_{0}$, respectively. As before,  $\left[\mathbb{H}_{m}^{\perp}\right]$, $\left[\mathbb{D}_{m}^{\parallel}\right]$, $\left[\mathbb{H}_{m}^{\parallel}\right]$, and $\left[\mathbb{D}_{m}^{\perp}\right]$ are column vectors containing the degrees of freedom of the modal fields.

We use the (discrete) Hodge star operator
~\cite{he2007differential,teixeira1999lattice,teixeira2014lattice,Gillette2011}
to convert the discrete Ampere's law from the dual mesh to the primal mesh. In this way,
\begin{flalign}
\left[\star_{\epsilon}\right]^{0\rightarrow0} \cdot \left[\mathbb{E}_{m}^{\perp}\right]^{n+1}
&=
\left[\star_{\epsilon}\right]^{0\rightarrow0} \cdot \left[\mathbb{E}_{m}^{\perp}\right]^{n}
\nonumber \\
&
+ \Delta t 
\left(  \left[\mathcal{D}_{\text{grad}}\right]^{{T}} \cdot \left[\star_{\mu^{-1}}\right]^{1\rightarrow1} \cdot  \left[\mathbb{B}_{m}^{\parallel}\right]^{n+\frac{1}{2}} \right),
\label{eq:DFL_primal_TM_pol}
\\
{\color{black}
\left[\star_{\epsilon}\right]^{1\rightarrow1} \cdot \left[\mathbb{E}_{m}^{\parallel}\right]^{n+1}
}
&
{\color{black}
=
\left[\star_{\epsilon}\right]^{1\rightarrow1} \cdot \left[\mathbb{E}_{m}^{\parallel}\right]^{n}
}
\nonumber \\
&
\!\!\!\!\!\!\!\!\!\!\!\!\!\!\!\!\!\!\!\!\!\!\!\!\!\!\!\!\!\!\!\!\!\!\!\!\!\!
{\color{black}
+ \Delta t 
\left( \left[\mathcal{D}_{\text{curl}}\right]^{{T}}\cdot  
\left[\star_{\mu^{-1}}\right]^{2\rightarrow2} \cdot \left[\mathbb{B}_{m}^{\perp}\right]^{n+\frac{1}{2}}
-\left|m\right| \left[\star_{\mu^{-1}}\right]^{1\rightarrow1} \cdot \left[\mathbb{B}_{m}^{\parallel}\right]^{n+\frac{1}{2}} \right),
}
\label{eq:DFL_primal_TE_pol}
\end{flalign}
where $\left[\tilde{\mathcal{D}}_{\text{curl}}\right]=\left[\mathcal{D}_{\text{grad}}\right]^{{T}}$, $\left[\tilde{\mathcal{D}}_{\text{grad}}\right]=\left[\mathcal{D}_{\text{curl}}\right]^{{T}}$ and the discrete Hodge matrix elements are given by~\cite{teixeira1999lattice,teixeira2014lattice,he2006geometric}
\begin{flalign}
&\!\!\!\!
\left[\star_{\epsilon}\right]_{J,j}^{1\rightarrow1}=\int_{\Omega}\left(\epsilon_{0}\rho\right) w_{J}^{(1)}\wedge \star\left(w_{j}^{(1)}\right)
=\underbrace{
\int_{\Omega}\left(\epsilon_{0}\rho\right)\mathbf{W}_{J}^{(1)} \cdot \mathbf{W}_{j}^{(1)} dV
}_{\text{vector proxy representation}},
\label{eq:DHG_eps_11}
\\
&\!\!\!\!
\left[\star_{\mu^{-1}}\right]_{K,k}^{2\rightarrow2}=\int_{\Omega}\left(\mu_{0}^{-1}\rho\right) w_{K}^{(2)}\wedge \star\left(w_{k}^{(2)}\right)
=\underbrace{
\int_{\Omega}\left(\mu_{0}^{-1}\rho\right)\mathbf{W}_{K}^{(2)} \cdot \mathbf{W}_{k}^{(2)} dV
}_{\text{vector proxy  rep.}},
\label{eq:DHG_mu_22}
\\
&\!\!\!\!
\left[\star_{\epsilon}\right]_{I,i}^{0\rightarrow0}=\int_{\Omega}\left(\epsilon_{0}\rho^{-1}\right) w_{I}^{(0)}\wedge \star\left(w_{i}^{(0)}\right)
=\underbrace{
\int_{\Omega}\left(\epsilon_{0}\rho^{-1}\right)\left[{\text{W}}_{I}^{(0)}\hat{\phi}\right] \cdot \left[{\text{W}}_{i}^{(0)}\hat{\phi}\right] dV
}_{\text{vector proxy  rep.}},
\label{eq:DHG_eps_00}
\\
&\!\!\!\!
{\color{black}
\left[\star_{\mu^{-1}}\right]_{J,j}^{1\rightarrow1}=\int_{\Omega}\left(\mu_{0}\rho\right)^{-1} w_{J}^{(\text{RWG})}\wedge \star \left(w_{j}^{(\text{RWG})}\right)
}
\nonumber \\
&~~~~~~~~~~~~~~~~~~~~~~~~~~~~~~~~~~~~~
{\color{black}
=\underbrace{
\int_{\Omega}\left(\mu_{0}\rho\right)^{-1}\left[\mathbf{W}_{J}^{(1)}\times\hat{\phi}\right] \cdot \left[\mathbf{W}_{j}^{(1)}\times\hat{\phi}\right] dV
}_{\text{vector proxy  rep.}},
}
\label{eq:DHG_mu_11}
\end{flalign}
where $\Omega$ is the (compact) spatial support of the Whitney forms, and the $\rho$, $\rho^{-1}$ factors result from the use of the TO in the mapping, as discussed before, where they enter as modifiers of constitutive properties rather than differential operator factors.
}
\color{black}{
The discrete Hodge matrices defined in (\ref{eq:DHG_eps_11}), (\ref{eq:DHG_mu_22}), (\ref{eq:DHG_eps_00}), and  (\ref{eq:DHG_mu_11}) are 
 instantiations of the (discrete) Galerkin-Hodge operator.  It should be emphasized that
the Galerkin-Hodge operator is not a natural consequence of DEC. The Galerkin-Hodge operator was originally proposed in \cite{Dodziuk}. It satisfies a number of built-in properties for stability in arbitrary simplicial meshes as discussed, for example, in references~\cite{teixeira2014lattice},\cite{Kettunen1},\cite{Bossjap},\cite{Kotiuga1}. In particular, the Galerkin-Hodge operator enforces standard local energy positivity~\cite{teixeira2013differential}.}

{\color{black}
The field updates in (\ref{eq:DFL_primal_TM_pol}) and (\ref{eq:DFL_primal_TE_pol}) call for sparse linear solvers due to the presence of the matrices $\left[\star_{\epsilon}\right]^{0\rightarrow0}$ and
$\left[\star_{\epsilon}\right]^{1\rightarrow1}$. From (\ref{eq:DHG_eps_11}) and (\ref{eq:DHG_eps_00}), it is seen that 
 $\left[\star_{\epsilon}\right]^{0\rightarrow0}$ and
$\left[\star_{\epsilon}\right]^{1\rightarrow1}$
are diagonally dominant and symmetric positive definite matrices; consequently, the linear solve can be performed very quickly. Nevertheless, this needs to be repeated at every time step. The linear solve can be obviated by computing a sparse approximate inverse (SPAI) of $\left[\star_{\epsilon}\right]^{0\rightarrow0}$ and
$\left[\star_{\epsilon}\right]^{1\rightarrow1}$ prior to the start of the time updating procedure. This strategy is discussed in~\cite{na2016local} and~\cite{kim2011parallel}. 
}
{\color{black}
The present algorithm is explicit and hence conditionally stable. The stability conditions are discussed in \ref{ap:stability_borfetd}.
}

{\color{black}
\subsection{Symmetry axis singularity treatment}
\begin{figure}
\centering
\subfloat[\label{fig:teppmc}]{\includegraphics[width=3in]{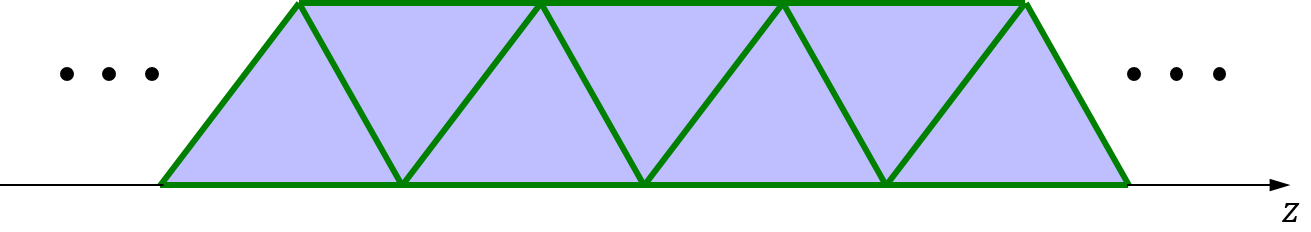}} 
\\
\subfloat[\label{fig:teppec}]{\includegraphics[width=3in]{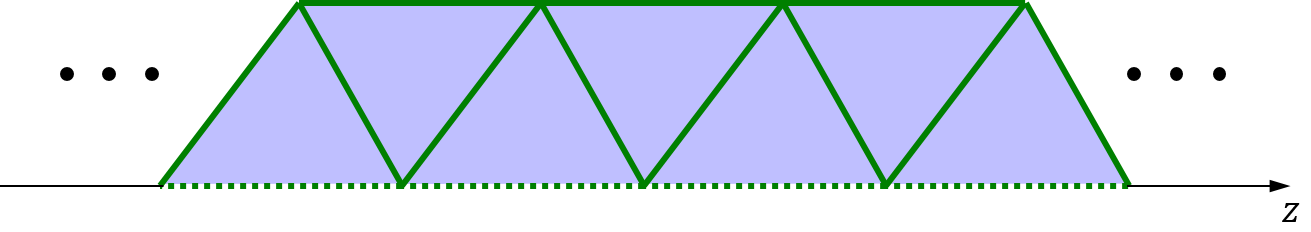}} 
\\
\subfloat[\label{fig:tmppmc}]{\includegraphics[width=3in]{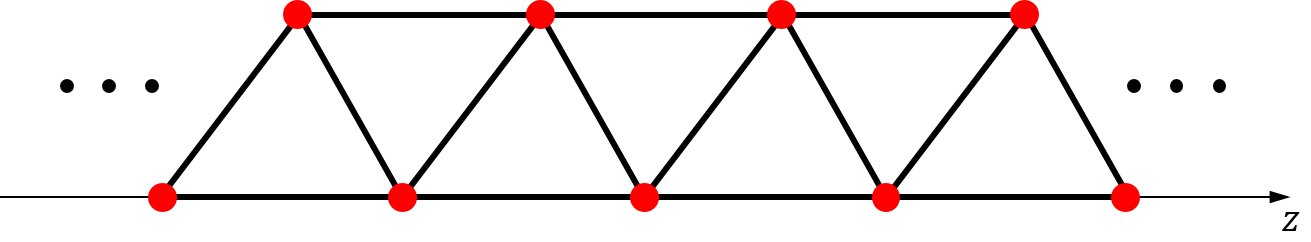}} 
\\
\subfloat[\label{fig:tmppec}]{\includegraphics[width=3in]{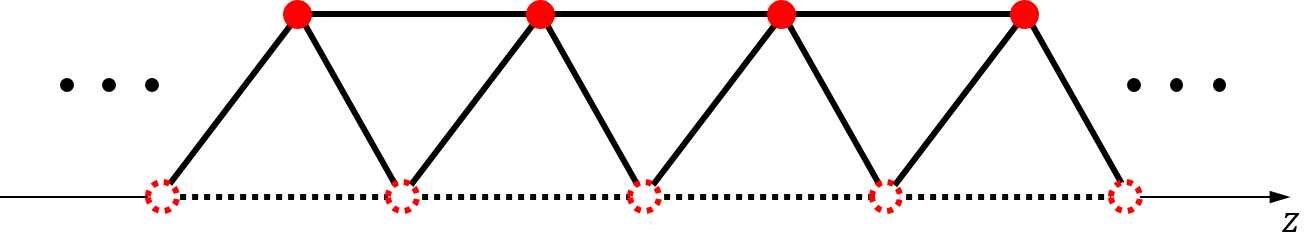}} 
\caption{Field boundary conditions on the primal mesh for the $\text{TE}^{\phi}$ field with (a) perfect magnetic conductor ($m=0$) and (b) perfect electric conductor ($m\neq0$) and for the $\text{TM}^{\phi}$ field with (c) perfect magnetic conductor ($m\neq0$) and (d) perfect electric conductor ($m=0$). Dashed lines indicate Dirichlet boundary condition, for example edges on the $z$ axis representing a perfect electric conductor boundary for $\text{TE}^{\phi}$ field in (b), or nodes on the $z$ axis representing a perfect electric conductor boundary for the $\text{TM}^{\phi}$ field in (d).}
\label{fig:fbc}
\end{figure}

For BOR problems where the line $\rho=0$ (symmetry axis) is part of the solution domain (for example, in hollow waveguides), it becomes necessary to treat the field behavior there by means of appropriate boundary
conditions.
The boundary conditions at $\rho=0$ are mode-dependent and should account for the cylindrical coordinate system singularity and the related degeneracy of the $\hat{\rho}$ and $\hat{\phi}$ unit vectors there.
When $m=0$, there is no field variation along azimuth and, in the absence of charges at $\rho=0$, both azimuthal and radial field components are zero at $\rho=0$.
On the other hand, the axial field component should be zero for $m\neq 0$~\cite{lebaric1989analysis} since the axial direction is invariant with respect to $\phi$ and a field dependency of the form {\color{black}$\cos \left(m\phi\right)$ or $\sin \left(m\phi\right)$} with $m \neq 0$ would imply a multivalued result at $\rho=0$ due to the coordinate degeneracy there.
As a result, when $m=0$, the boundary $\rho=0$ can be represented as a  perfect electric conductor for the $\text{TE}^{\phi}$ field and as a perfect magnetic conductor for the $\text{TM}^{\phi}$ field.
Conversely, when $m\neq 0$, the $\rho=0$ boundary can be represented as a perfect magnetic conductor  for the $\text{TE}^{\phi}$ field and as a perfect electric conductor for the  $\text{TM}^{\phi}$ field.
A homogeneous Neumann boundary condition for the electric field can be used to represent the perfect magnetic conductor case and a homogeneous Dirichlet boundary condition for the perfect electric conductor case.
Implementation of such boundary conditions on the primal mesh is illustrated in Fig.~\ref{fig:fbc}.
Dashed lines in Fig. \ref{fig:teppec} and \ref{fig:tmppec} denote the Dirichlet boundary implementation: along the $z$ axis, the perfect electric conductor condition is enforced on grid edges for the $\text{TE}^{\phi}$ case and on grid nodes for the $\text{TM}^{\phi}$ case.
Likewise, Fig. \ref{fig:teppmc} and \ref{fig:tmppmc} illustrate application of the Neumann boundary condition:
along the $z$ axis, the perfect magnetic conductor condition is enforced on grid edges for the $\text{TE}^{\phi}$ case and on grid nodes for the $\text{TM}^{\phi}$ case.

Using the boundary conditions described above, the present FETD-BOR Maxwell solver does not require any modifications in the basis functions on the grid cells adjacent to the $z$ axis, unlike prior FE-BOR Maxwell solvers.
}

{\color{black}
\section{Numerical Examples}
In order to validate present FETD-BOR Maxwell solver, we first consider a cylindrical cavity and compare the resonance frequency results to the analytical predictions.
Then, we illustrate two practical examples of devices based on BOR geometries: logging-while-drilling sensors used for Earth formation resistivity profiling in geophysical exploration and 
relativistic BWO for high-power microwave applications.

\subsection{Cylindrical cavity}
We simulate the eigenfrequencies of a hollow cylindrical cavity with metallic walls using the present FETD-BOR Maxwell solver,  and compare the results to analytic predictions.
The cavity has radius $a=0.5$ m and height $h=1$ m, as depicted in Fig. \ref{fig:cyl_cav_geom}.
Magnetic and electric dipole current sources $\mathbf{M}\left(\mathbf{r},t\right)$ and $\mathbf{J}\left(\mathbf{r},t\right)$ oriented along $\phi$ and excited by broadband Gaussian-modulated pulses are placed at arbitrary locations inside the cavity $\mathbf{r}_{s}=\left(\rho_{s},\phi_{s},z_{s}\right)$, so that
\begin{flalign}
&\mathbf{M}\left(\mathbf{r},t\right)~,~\mathbf{J}\left(\mathbf{r},t\right)=
\hat{\phi} \, G(t) \,\delta\left( \mathbf{r}-\mathbf{r}_{s} \right)=
\nonumber\\
&=\hat{\phi} \, G(t) \, \delta\left( \mathbf{r}^{\parallel}-\mathbf{r}_{s}^{\parallel} \right)\left[\pi+2\pi\sum_{m=1}^{M_{\phi}}\cos{\left[m\left(\phi-\phi_{s}\right)\right]}\right]
\label{eqn:MJ_source}
\end{flalign}
where $G(t)=e^{-\left[\left({t-t_{g}}\right)/\left({2\sigma_{g}}\right)\right]^{2}}\sin{\left[2\pi f_{g}\left(t-t_{g}\right)\right]}$ with $t_{g}=20$ ns, $\sigma_{g}=1.9$ ns, and $f_{g}=300$ MHz, and $\mathbf{r}^{\parallel}=\rho\hat{\rho}+z\hat{z}$. 
We use Fourier series expansion to describe $\delta\left( \phi-\phi_{s} \right)$ in (\ref{eqn:MJ_source}) in order to match the modal field expansion used before.
A total of four dipole sources (electric and magnetic currents) are used to excite a rich gamut of eigenmodes, as illustrated in Fig. \ref{fig:cyl_cav_geom}. 
\begin{figure}
    \centering
    \includegraphics[width=2.75in]{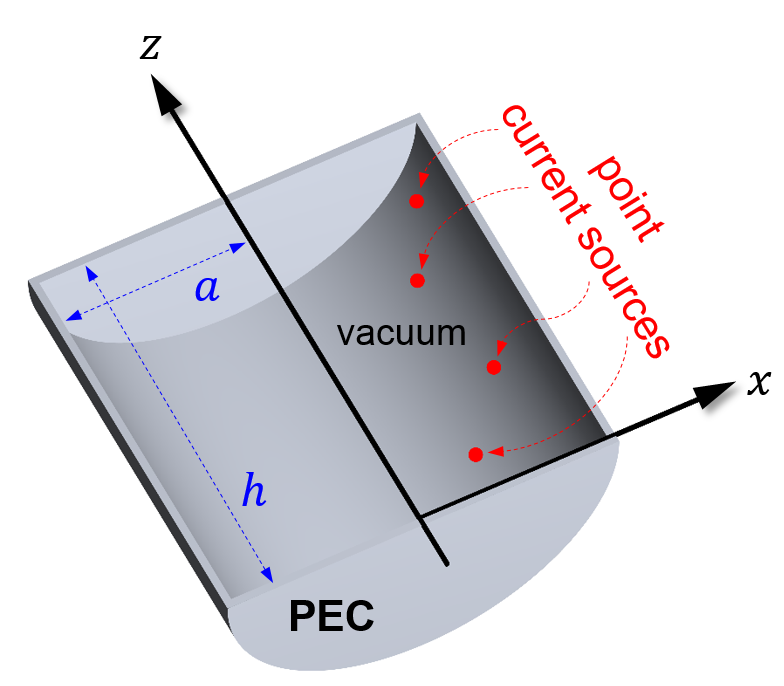}
	\caption{Schematic view of the simulated cylindrical cavity with perfect electric conductor (PEC) walls. The cavity dimensions are $a=0.5$ m and $h=1$ m.}
\label{fig:cyl_cav_geom}
\end{figure}
The meridian plane of the cylindrical cavity is discretized by an unstructured grid with $4,045$ nodes, $11,939$ edges, and $7,895$ faces (seen as the $\rho z$ plane for $\phi=180^{o}$ in Fig. \ref{fig:cyl_cavity_field_E}).
The metallic boundaries are treated as perfect electric conductors.
{\color{black}In this case, the maximum azimuthal modal order $M_{\phi}$ was set equal to $4$ to investigate the field solution up to this order.
Higher order modes can be included by simply increasing $M_{\phi}$. This is straightforward since azimuthal modal fields with different orders are orthogonal to each other.
From the stability analysis in \ref{ap:stability_borfetd}, the maximum time-step intervals for various cases are presented in Table~\ref{tab:cavity_stability}.
\begin{table}
\caption{Maximum time-step intervals for various cases in the simulation of cylindrical metallic cavity.}  
\centering
  \begin{tabular}{l*{6}{c}}
    \toprule
    & \multicolumn{2}{c}{$m=0$}  & \multicolumn{4}{c}{$m\neq0$} \\
    \cmidrule(lr){2-3}\cmidrule(lr){4-7}
    & \multicolumn{1}{c}{$\text{TE}^{\phi}$-pol.} & \multicolumn{1}{c}{$\text{TM}^{\phi}$-pol.} & \multicolumn{1}{c}{$m=1$} & \multicolumn{1}{c}{$m=2$} & \multicolumn{1}{c}{$m=3$} & \multicolumn{1}{c}{$m=4$} \\
    \midrule
    \addlinespace
    $\Delta t_{\text{max}}$ [ps]  &  10.009    &  10.249    &  10.009   &  6.4792  &  4.5545   &  3.4843      \\
    \bottomrule
  \end{tabular}
\label{tab:cavity_stability}
\end{table}
Here we chose $\Delta {t}=1$ ps for the simulations and used a total of $1 \times10^{7}$ time steps to provide sufficiently narrow resonance peaks.} 
By recording the time history of the electric field values at arbitrary locations inside the cavity and performing a Fourier transform, we obtain the eigenfrequencies as peaks in the Fourier spectrum.
Fig. \ref{fig:cyl_cav_spectrum} shows the normalized spectral amplitude as a function of frequency.
The black solid line is the result obtained by using present FETD-BOR Maxwell solver. The red dashed and blue solid lines indicate analytic predictions for the eigenfrequencies of the $\text{TE}_{mnp}$ and $\text{TM}_{mnp}$ modes in this cavity, respectively.
The analytic expressions for the eigenfrequencies are given by
\begin{flalign}
f_{\text{TE}_{mnp}}&=\frac{2 c}{\pi}\sqrt{{\chi'}_{mn}^{2}+\left({\frac{p \pi}{h}}\right)^{2}},
\nonumber
\\&\text{for}~m=0,1,...,~n=1,2,...,~p=1,2,...~,
\\
f_{\text{TM}_{mnp}}&=\frac{2 c}{\pi}\sqrt{\chi_{mn}^{2}+\left({\frac{p \pi}{h}}\right)^{2}},
\nonumber
\\&\text{for}~m=0,1,...,~n=1,2,...,~p=0,1,...~,
\end{flalign}
where $c$ is speed of light, $\chi_{mn}$ and ${\chi'}_{mn}$ are the roots of the equations $J_{m}\left(a\chi_{mn}\right)=0$ and ${J'}_{m}\left(a{\chi'}_{mn}\right)=0$, respectively, with
$J_{m}\left(\cdot \right)$  being the Bessel function of first kind and ${J'}_{m}\left(\cdot \right)$  its derivative with respect to the argument.
\begin{figure}
    \centering
    \includegraphics[width=3.4in]{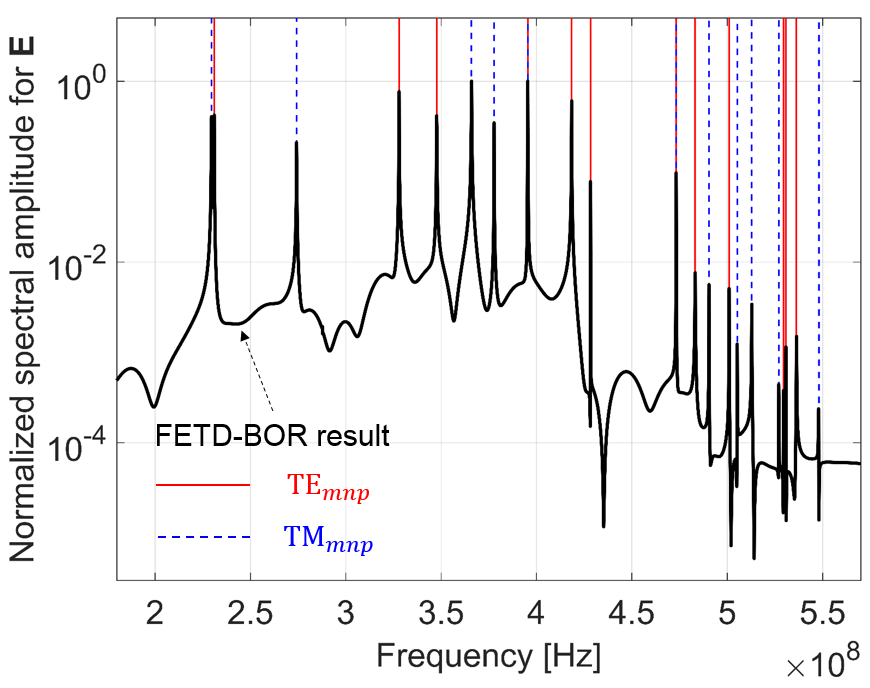}
	\caption{Normalized spectral amplitude for $\mathbf{E}$, showing the eigenfrequencies of the cavity. Black solid lines correspond to the present FETD-BOR result. Red solid and blue dashed lines are analytic predictions for the $\text{TE}_{mnp}$ and $\text{TM}_{mnp}$ eigenfrequencies, respectively.}
\label{fig:cyl_cav_spectrum}
\end{figure}
It is clear from Fig. \ref{fig:cyl_cav_spectrum} that there is a great agreement between the simulated and analytic eigenfrequencies. Table \ref{tab:spectrum_cavity} shows the relative error between the simulated $f_s$ and analytical $f_a$ frequencies. The relative error is below $0.03~\%$ in all cases, indicating the accuracy of the proposed field solver.
\begin{table} [t]
\caption{Eigenfrequencies for the cylindrical cavity and normalized errors between numerical and analytic results.}  
\centering
{\normalsize 
\begin{tabular}{ccc}
\toprule
\addlinespace
Resonant modes & $f_{a}$ [MHz] & $\left|f_{a}-f_{s}\right|/f_{a}\times 100$ [\%] 
\\
\midrule
$\text{TM}_{010}$   	& $229.6369$ 						& $1.1854 \times10^{-2}$\\
$\text{TE}_{111}$   	& $231.1104$ 						& $8.0278 \times10^{-4}$\\
$\text{TM}_{011}$   	& $274.2865$ 						& $2.4558  \times10^{-2}$\\
$\text{TE}_{211}$   	& $327.9619$ 						& $1.0503  \times10^{-2}$\\
$\text{TE}_{112}$   	& $347.7241$ 						& $1.7614  \times10^{-2}$\\
$\text{TM}_{110}$   	& $365.8931$ 						& $2.8110  \times10^{-2}$\\
$\text{TM}_{012}$   	& $377.8003$ 						& $7.0851  \times10^{-3}$\\
$\text{TE}_{011}\text{,}~\text{TM}_{111}$ & $395.4463$ & $1.5709   \times10^{-3}$\\
$\text{TE}_{212}$   	& $418.4005$ 						& $9.3816    \times10^{-3}$\\
$\text{TE}_{311}$   	& $428.3025$ 						& $6.0946    \times10^{-3}$\\
$\text{TE}_{012}\text{,}~\text{TM}_{112}$ & $473.1572$ & $1.2629   \times10^{-2}$\\
$\text{TE}_{113}$   	& $483.1273$ 						& $4.4680   \times10^{-3}$\\
$\text{TM}_{210}$   	& $490.4134$ 						& $5.2154  \times10^{-3}$\\
$\text{TE}_{312}$   	& $500.9421$ 						& $1.0443   \times10^{-2}$\\
$\text{TM}_{013}$   	& $505.2060$ 						& $2.0998 \times10^{-3}$\\
$\text{TM}_{211}$   	& $512.8404$ 						& $8.3352  \times10^{-3}$\\
$\text{TM}_{020}$   	& $527.1202$ 						& $2.3910    \times10^{-2}$\\
$\text{TE}_{411}$   	& $529.4750$ 						& $6.8899   \times10^{-3}$\\
$\text{TE}_{121}$   	& $530.7481$ 						& $2.5411   \times10^{-2}$\\
$\text{TE}_{213}$   	& $536.2453$ 						& $5.4133 \times10^{-3}$\\
$\text{TM}_{021}$   	& $548.0472$ 						& $2.9989 \times10^{-2}$\\
\bottomrule
\end{tabular}
}
\label{tab:spectrum_cavity}
\end{table}

To illustrate the field behavior, Figs.~\ref{fig:cyl_cavity_field_E}  and ~\ref{fig:cyl_cavity_field_B} show snapshots for electric field intensity and magnetic flux density distribution inside the cavity on four $\rho z$ planes with $\phi=0^{o}$, $\phi=90^{o}$, $180^{o}$, $270^{o}$ and two $\rho\phi$ planes with $z=0.2$ m and $0.8$ m, at four time instants: $1.0024~\mu\text{s}$, $1.0028~\mu\text{s}$, $1.0032~\mu\text{s}$, and $1.0036~\mu\text{s}$.
Due to the location of the dipole sources, the transient fields produced include many eigenmodes, and are basically asymmetric.
It can be seen that the (tangential or normal) boundary conditions on the outer perfect electric conductor walls for electric field intensity and magnetic flux density are well satisfied.
Moreover, the correct field distribution along the symmetry axis is well reproduced by the chosen boundary conditions at $\rho=0$, without any spurious artifacts.
\begin{figure*}
\centering
\subfloat[\label{fig:}]{\includegraphics[width=2.5in]{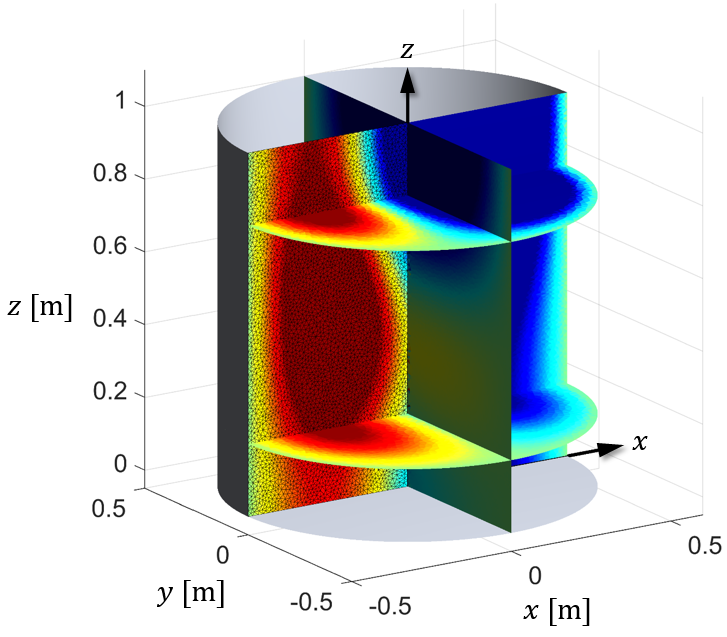}} 
\subfloat[\label{fig:}]{\includegraphics[width=2.5in]{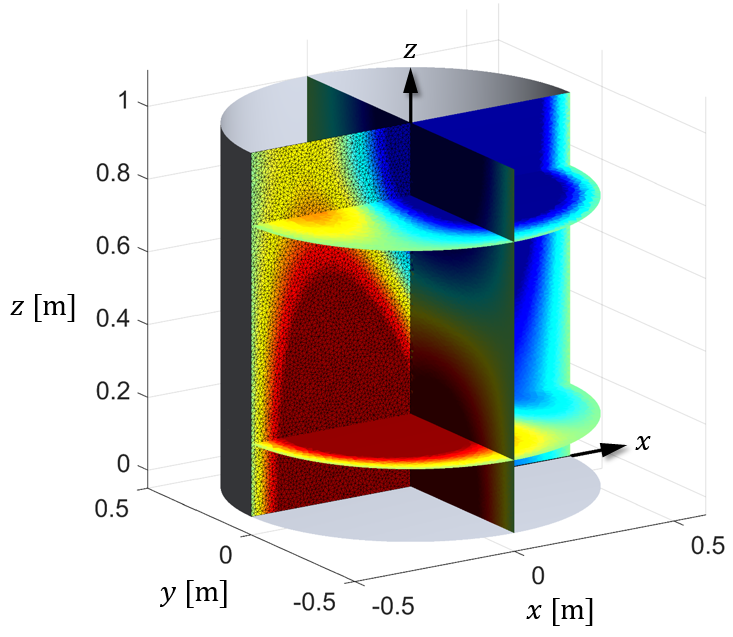}} 
\\
\subfloat[\label{fig:}]{\includegraphics[width=2.5in]{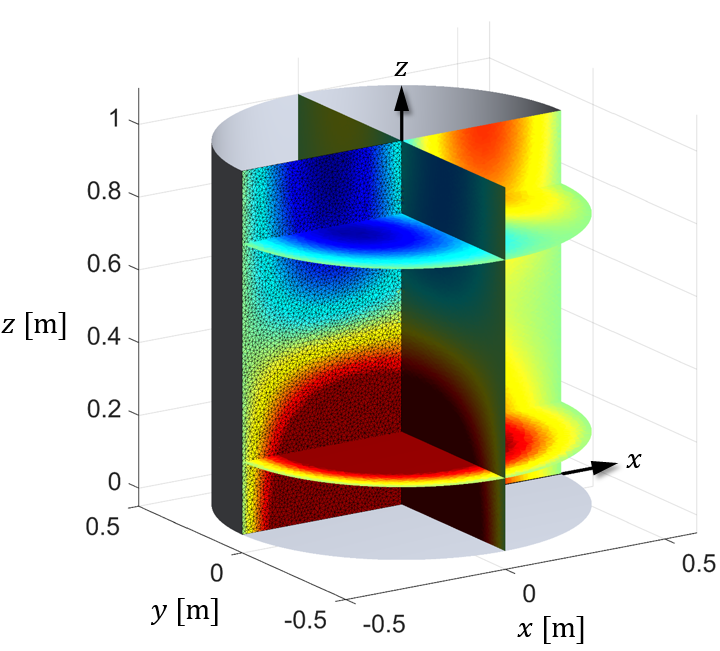}} 
\subfloat[\label{fig:}]{\includegraphics[width=2.5in]{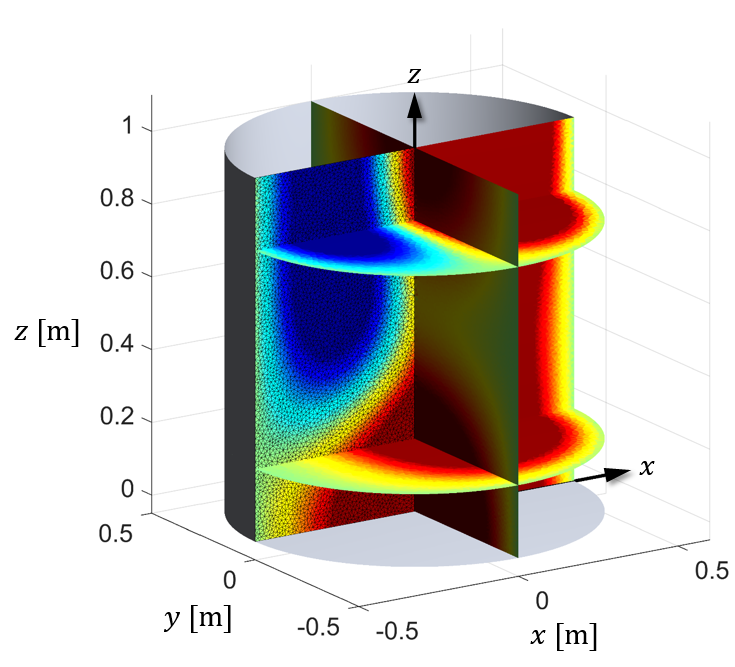}} 
\caption{Transient snapshots for $E_{z}$ inside the cylindrical cavity at (a) $1.0024~[\mu \text{s}]$, (b) $1.0028~[\mu \text{s}]$, (c) $1.0032~[\mu \text{s}]$, and (d) $1.0036~[\mu \text{s}]$.}
\label{fig:cyl_cavity_field_E}
\end{figure*}
\begin{figure*}
\centering
\subfloat[\label{fig:}]{\includegraphics[width=2.5in]{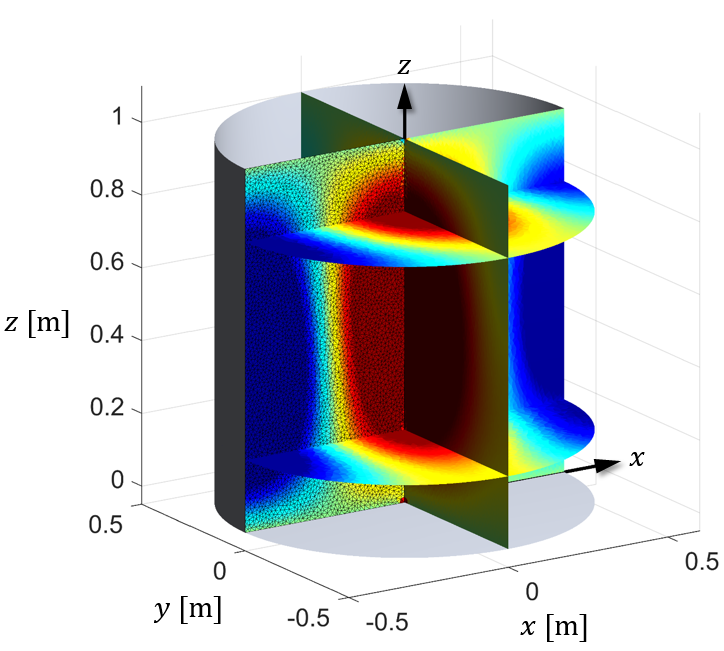}}
\subfloat[\label{fig:}]{\includegraphics[width=2.5in]{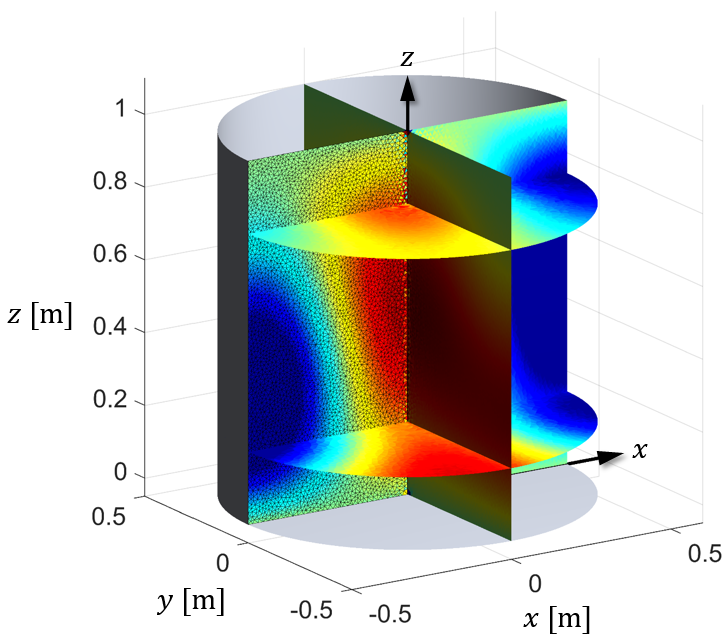}}
\\
\subfloat[\label{fig:}]{\includegraphics[width=2.5in]{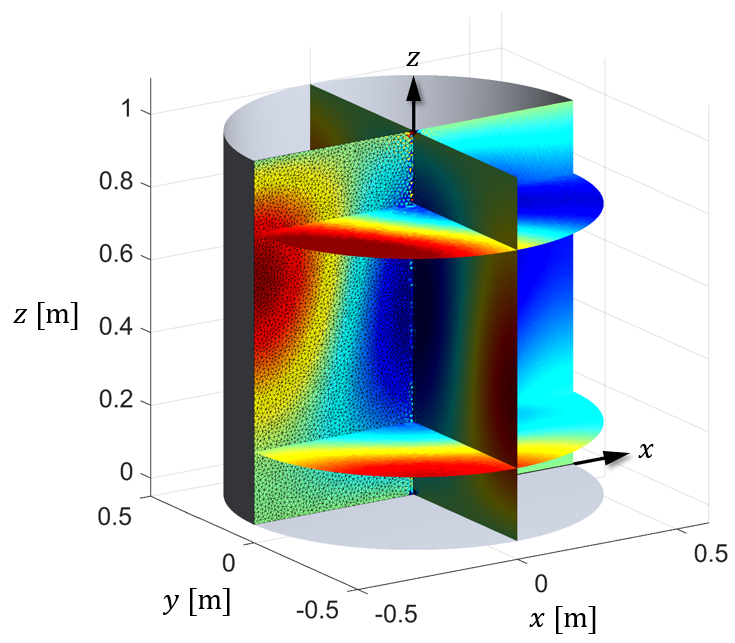}}
\subfloat[\label{fig:}]{\includegraphics[width=2.5in]{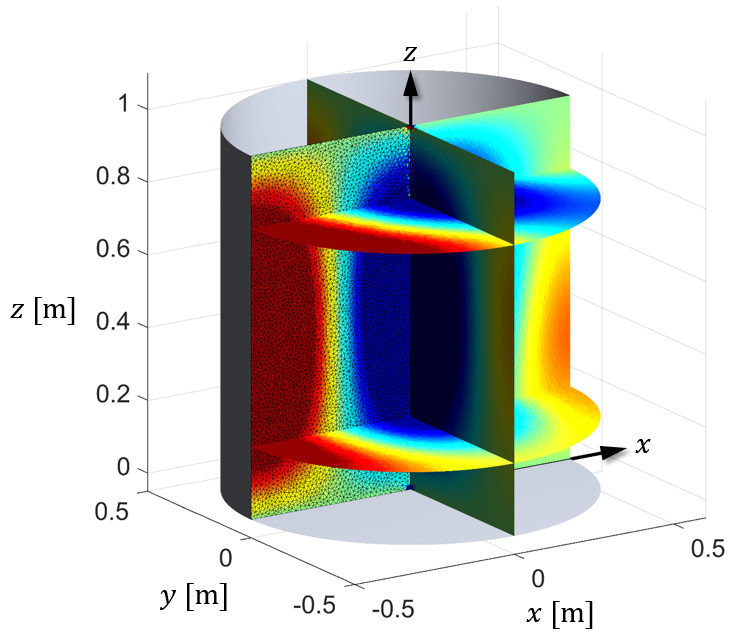}}
\caption{Transient snapshots for $B_{z}$ inside the cylindrical cavity at (a) $1.0024~[\mu \text{s}]$, (b) $1.0028~[\mu \text{s}]$, (c) $1.0032~[\mu \text{s}]$, and (d) $1.0036~[\mu \text{s}]$.}
\label{fig:cyl_cavity_field_B}
\end{figure*}

\subsection{Logging-while-drilling sensor simulation}
Logging-while-drilling sensors have BOR geometries and are routinely used for hydrocarbon exploration~\cite{novo2008comparison, novo2010three,hong2017novel, yang2017stable, fang2017through, hue2005three}. 
As the drilling process is performed, these sensors record logs obtained by the measurements of fields produced by loop (multi-coil) antennas present in the sensor and reflected from the surrounding geological formation.
Logging-while-drilling sensors are typically equipped with a series of transmitter and receiver loop antennas that are wrapped around the outer diameter of a metallic mandrel attached to the bit drill~\cite{li2016investigation, zhang2001simulation,  novo2011comparative, lee2012numerical, liu2012analysis, rosa2017robust}.
Fields produced by the transmitter coil(s) interact with the adjacent well-bore environment and are detected by a pair (or more) of receiver coils along the logging-while-drilling sensor at same axial distance from the transmitter(s). 
Two types of measurements are typically used to determine the resistivity profiles of the adjacent formation. 
The first is the amplitude ratio (AR) between the electromotive force (e.m.f.) excited at the two receiver coils and the second is their phase difference (PD).
In this section, we consider a prototypical concentric logging-while-drilling sensor generating a $\text{TM}^{\phi}$ field distribution in the formation with $m=0$\footnote{Not only the geometry but also the field excitation is axisymmetric in this case.}.
The logging-while-drilling sensor depicted in Fig. \ref{fig:LWD_configuration} consists of a metallic cylindrical mandrel modeled as a perfect electric conductor inside a concentric cylindrical borehole. Three loop antennas are used: one as transmitter and two as receivers. The borehole created by the drilling process is filled with a lubricant fluid (mud). 
The three coil antennas are moving downward in tandem as the drilling process occur.
\begin{figure}[t]
    \centering
    \includegraphics[width=2.3in]{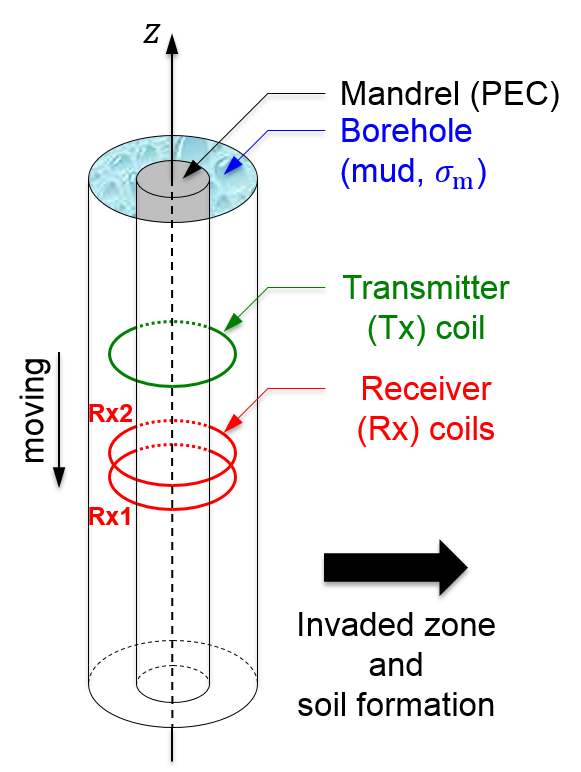}
	\caption{Logging-while-drilling sensor problem geometry (from inner to outer features): metallic mandrel, transmit (Tx) and receive (Rx) coil antennas, mud-filled borehole, and adjacent geological formation.}
\label{fig:LWD_configuration}
\end{figure}

We consider two scenarios for the adjacent Earth formation, as shown in Fig. \ref{fig:two_scenarios}.
In the first scenario, the borehole is filled with a low conductive (oil-based) fluid (mud) having $\sigma =0.0005$ S/m and surrounded by geological formations with different conductivities.
We compute the AR and PD as a function of the formation conductivity.
In the second scenario, the borehole is filled with a high conductive (water-based) fluid having $\sigma =2 $ S/m,
and the formation has three horizontal layers with different conductivities as shown. We compute the AR and PD as the set of coil antennas (sensor) moves downward.
In both cases, the relative permittivity and permeability are assumed equal to one everywhere, and the transmitter coil radiates a $2$ MHz signal. In the time domain, this is implemented through a current signal along the transmitter coil given by $I_{\text{Tx}}(t)=r(t)\sin{(\omega t)}$, where  
\begin{eqnarray}
r(t) = 
\begin{dcases}
0, & t < 0 \\
0.5\left[1-\cos{\left(\frac{\omega t}{2 \alpha}\right)}\right], & 0 \leqslant t < \alpha T\\
1, & t \geqslant \alpha T,
\end{dcases}
\end{eqnarray}
is a raised-cosine ramp function, $T=2\pi/\omega$ is the signal period, and $\alpha$ is the number of sine wave cycles during the ramp duration $\alpha T$.
The use of ramp function mitigates high frequency components otherwise produced by an abrupt turn-on at $t=0$, and yields faster convergence of AR and PD (after approximately one time period $T$)~\cite{hue2005three}. We choose $\alpha=0.5$ to yield a continuous first-order derivative and no DC (zero-frequency) component for the signal.
From the time-domain signals computed at the two receivers, we extract the corresponding phases $\theta$ and amplitudes $A$ using
\begin{flalign}
&
\theta=\tan^{-1}{\left( \frac{q_{2}\sin{(\omega t_{1})}-q_{1}\sin{(\omega t_{2})}}{q_{1}\cos{(\omega t_{2})}-q_{2}\cos{(\omega t_{1})}} \right)}, \\
&
A=\left|\frac{q_{1}}{\sin{(\omega t_{1}+\theta)}}\right|,
\end{flalign}
where $q_{1}$ and $q_{2}$ are signals computed at times $t_{1}$ and $t_{2}$, respectively~\cite{hue2005three}. 
Next, the AR and PD are calculated as
\begin{flalign}
&
\text{AR}=A_{ \text{Rx}_{2} }/A_{ \text{Rx}_{1} },\\
&
\text{PD}=\theta_{ \text{Rx}_{2} }-\theta_{ \text{Rx}_{1} }.
\end{flalign}
\begin{figure}[t]
     \centering	
	\subfloat[\label{fig:w_1}]{
     \includegraphics[width=1.5in]{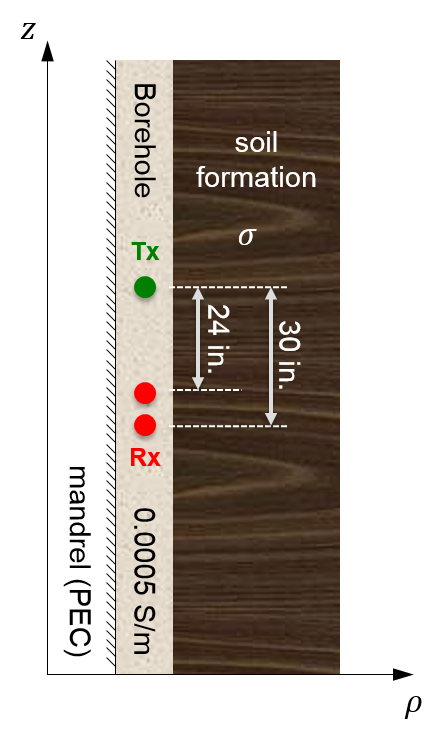}
	}
	\subfloat[\label{fig:w_2}]{
     \includegraphics[width=1.8in]{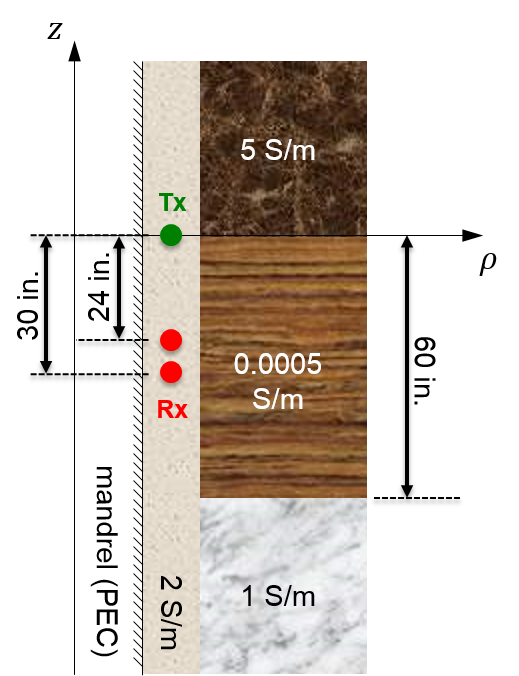}
	}
	\caption{Logging-while-drilling sensor responses. (a) First scenario: the conductivity of the adjacent geological formation is varied. (b) Second scenario: the sensor moves downward through a borehole surrounded by a geological formation with three horizontal layers.}
\label{fig:two_scenarios}
\end{figure}
The azimuthal electric current along the transmitter coil is modeled as a nodal current density on the meridian plane and the metallic mandrel is regarded as perfect electric conductor.
The FE domain is truncated by a PML to mimic an open domain. We use 8 layers for the PML to yield a reflectance below $-50$ dB \cite{donderici2008mixed}.

Fig. \ref{fig:arpdsf} shows results for the behavior of AR and PD versus the conductivity on a homogeneous formation. The results are compared against previous results obtained by the finite-difference time-domain (FDTD) and the numerical mode matching (NMM) methods~\cite{hue2005three}. There is excellent agreement between the results.
\begin{figure}[t]
     \centering	
	\subfloat[\label{fig:AR_sf}]{
     \includegraphics[width=3in]{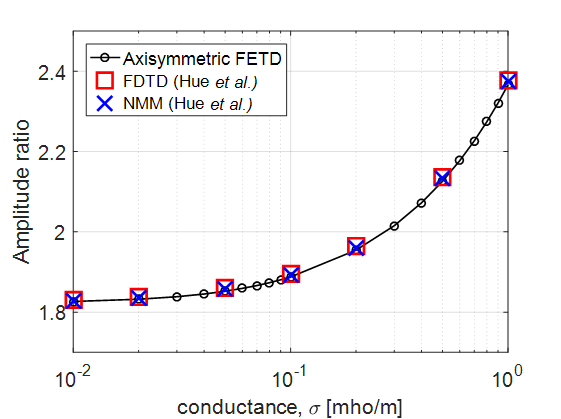}
	}
\\
	\subfloat[\label{fig:PD_sf}]{
     \includegraphics[width=3in]{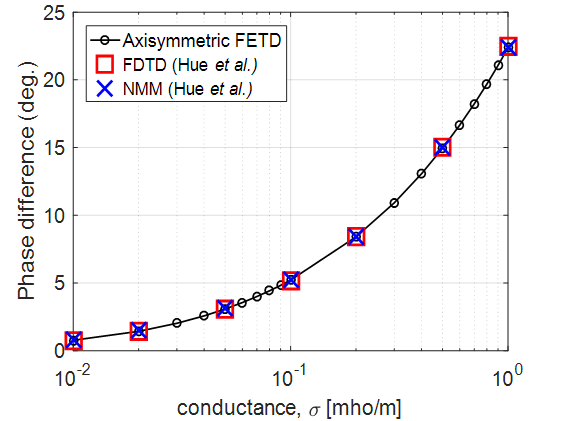}
	}
	\caption{Computed (a) AR and (b) PD (in deg.) by a logging-while-drilling sensor surrounded by homogeneous geological formations with different conductivities. This corresponds to the first scenario in Fig. \ref{fig:two_scenarios}. The results from the present algorithm are compared against FDTD and NMM results~\cite{hue2005three} (see more details in the main text).}
\label{fig:arpdsf}
\end{figure}
\begin{figure}
    \centering
    \includegraphics[width=3in]{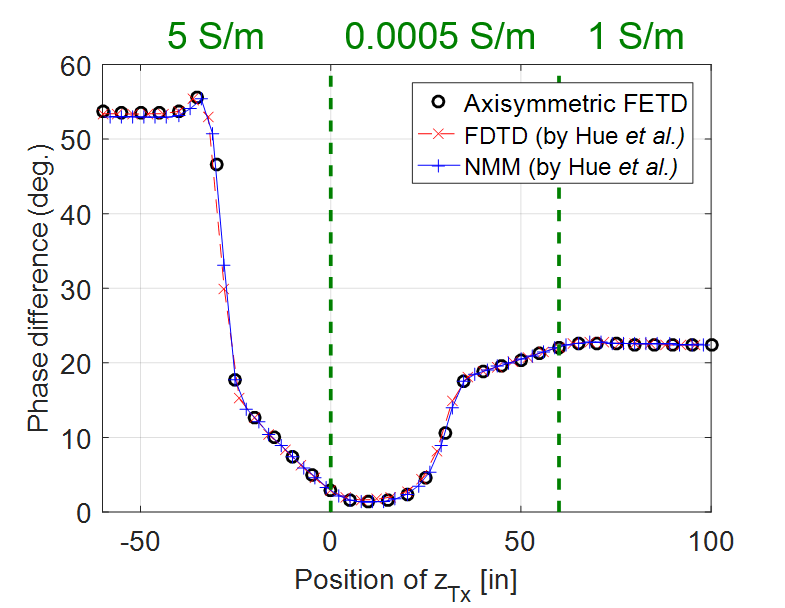}
	\caption{Computed PD (deg.) between the two receivers of the logging-while-drilling sensor versus the $z$ position of the transmitter coil antenna. This corresponds to the second scenario in Fig. \ref{fig:two_scenarios}.  The results from the present algorithm are compared against FDTD and NMM results~\cite{hue2005three}  (see more details in the main text).}
\label{fig:pdss}
\end{figure}
\begin{figure*}
\centering
	\subfloat[\label{fig:fd_case_1_ss}]{     
    \includegraphics[width=2.55in]{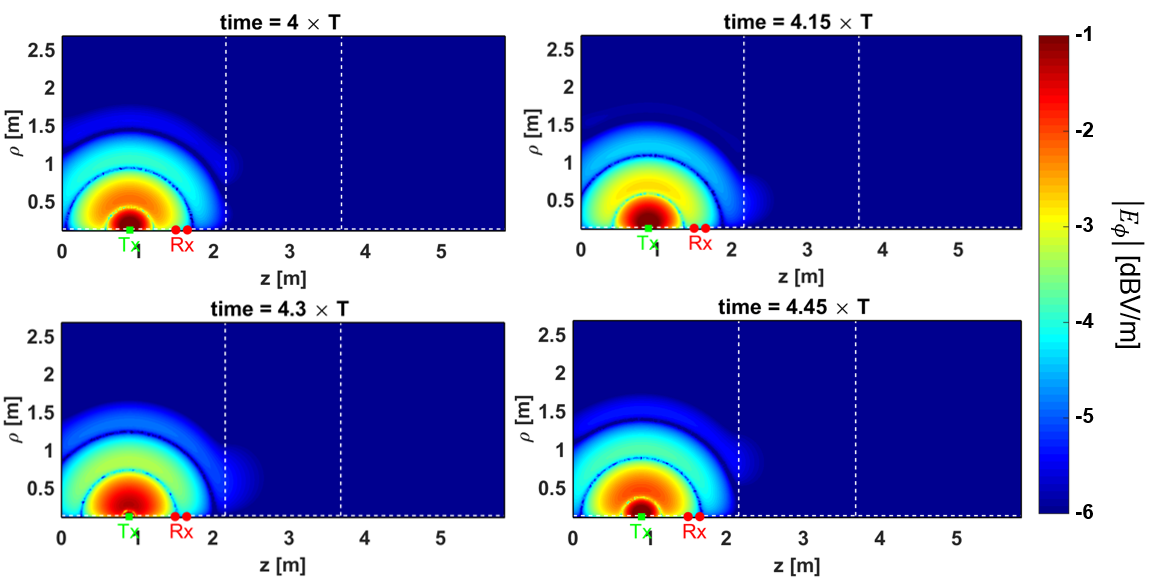}
}
	\subfloat[\label{fig:fd_case_2_ss}]{     
    \includegraphics[width=2.55in]{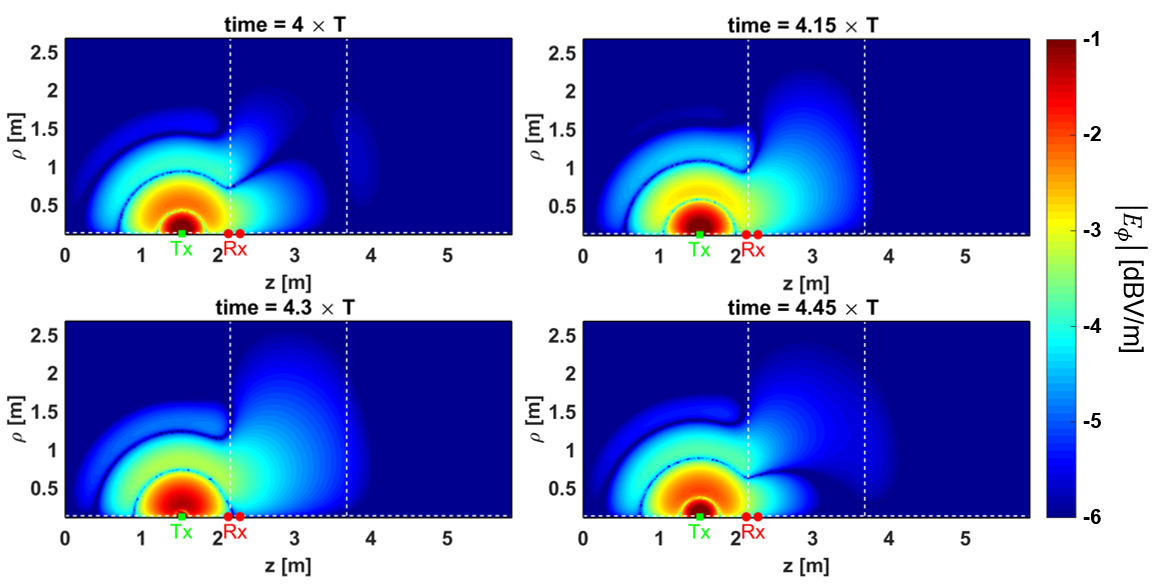}
}
\\
	\subfloat[\label{fig:fd_case_3_ss}]{     
      \includegraphics[width=2.55in]{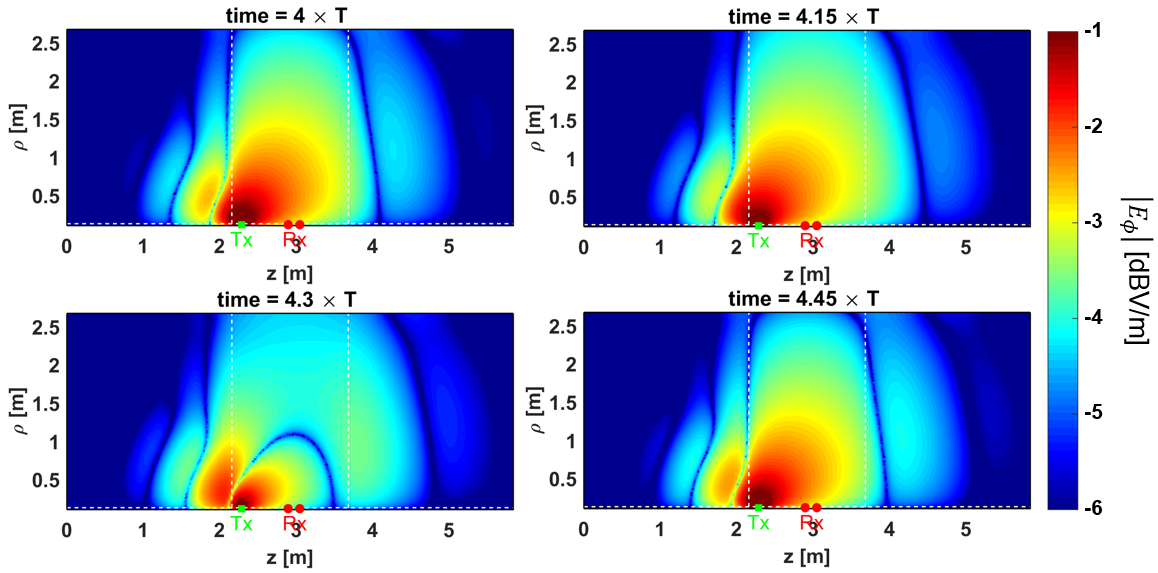}
	}
	\subfloat[\label{fig:fd_case_4_ss}]{     
    \includegraphics[width=2.55in]{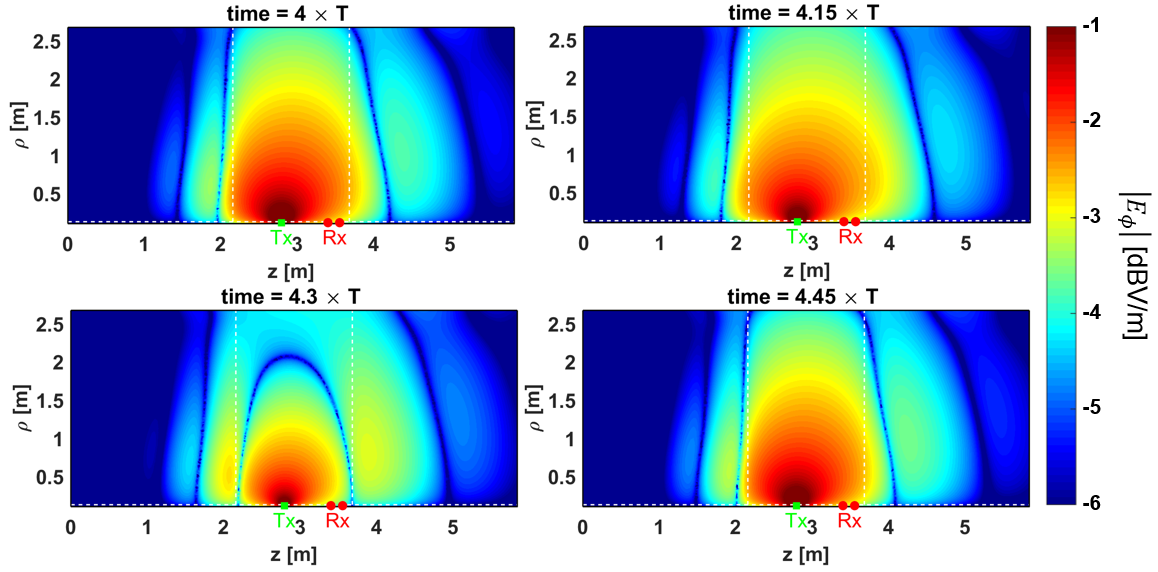}
}
\\
	\subfloat[\label{fig:fd_case_7_ss}]{     
    \includegraphics[width=2.55in]{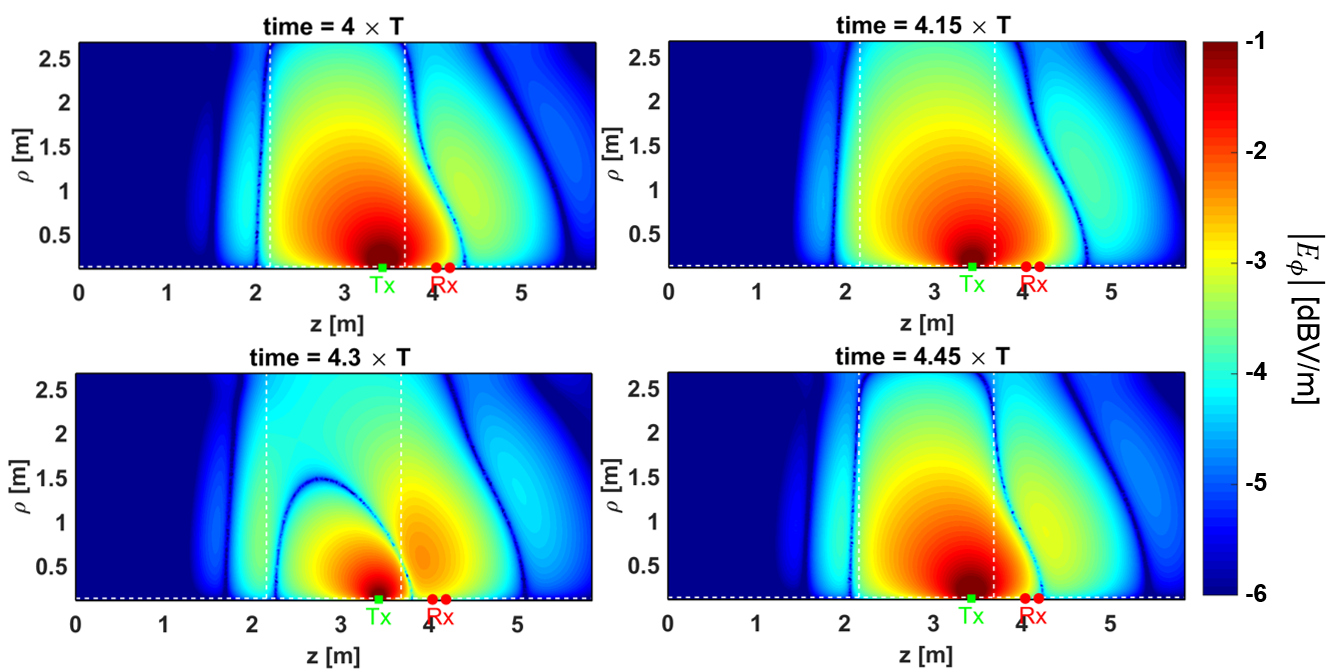}
}
	\subfloat[\label{fig:fd_case_6_ss}]{     
    \includegraphics[width=2.55in]{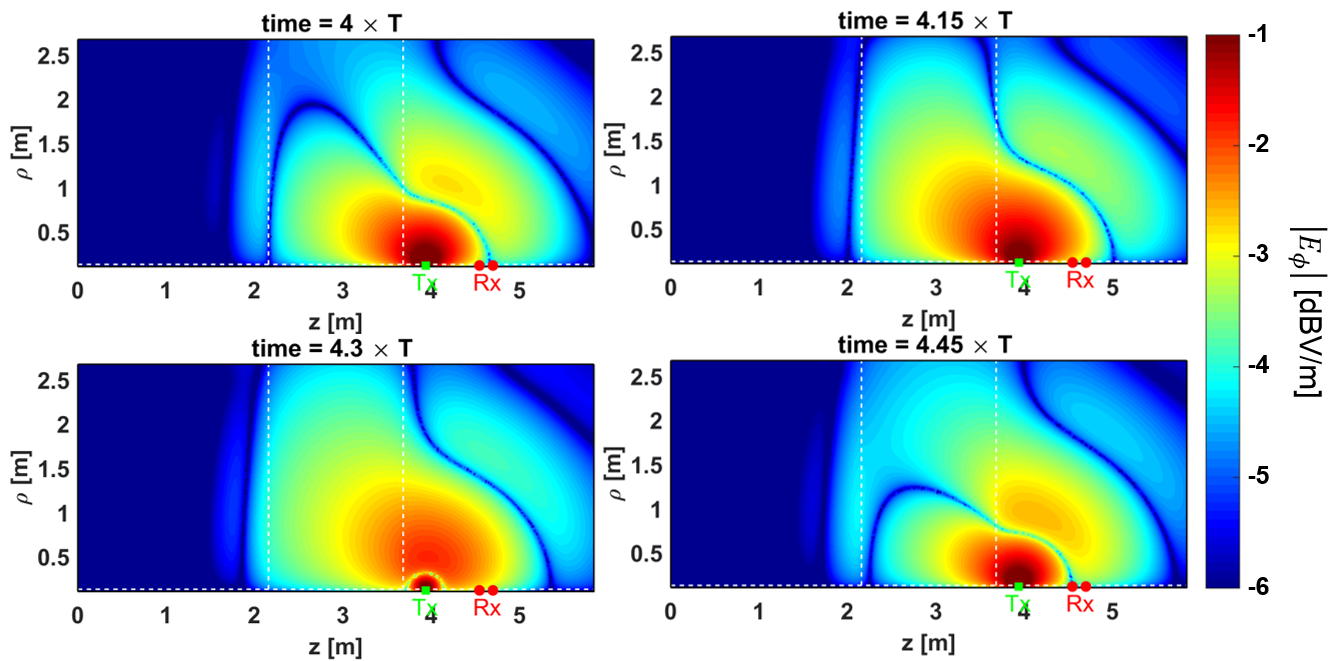}
}
	\caption{Electric field distribution during the half period for $z_{\text{Tx}}=$ (a) $-50$ inch, (b) $-25$ inch, (c) $5$ inch, (d) $25$ inch, (e) $50$, and (f) $70$ inch. Note that $z_{\text{Tx}}=0$ at the interface between first ($5$ S/m) and second ($0.0005$ S/m) formations.}
\label{fig:fd_case_ss}
\end{figure*}
Results for the second scenario are shown in Fig. \ref{fig:pdss}, where PD is plotted as a function of the $z$ position of the transmitter, $z_{\text{Tx}}$, and compared against previous results obtained by the FDTD and NMM methods~\cite{hue2005three}. Again, an excellent agreement is obtained.
As expected, the PD is higher when the coil antennas are within high attenuation (high conductivity) layer and vice versa.
 The conductance profile and the corresponding axial extension of each formation is shown in green color in Fig. \ref{fig:pdss}.
Fig. \ref{fig:fd_case_1_ss}$-$Fig. \ref{fig:fd_case_6_ss} show snapshots of the electric field distributions for different $z_{\text{Tx}}$ to illustrate the field behavior.

\subsection{Backward-wave oscillator (BWO) in the relativistic regime}
\begin{figure}
\centering
\includegraphics[width=4.3in]{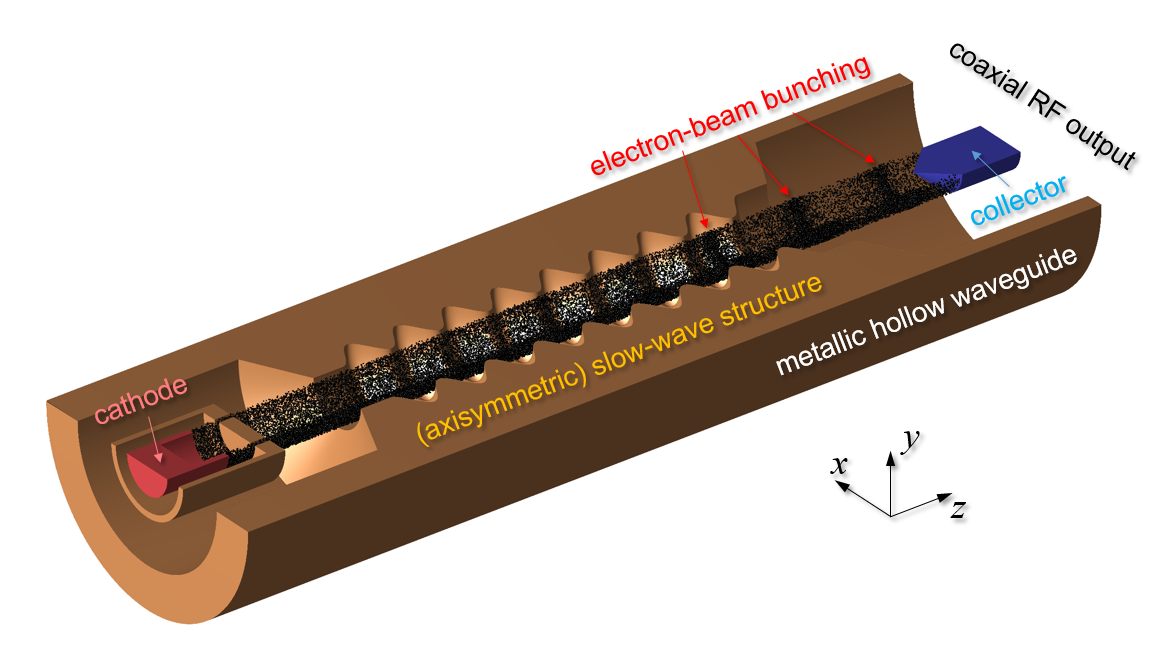}
\caption{Relativistic backward-wave oscillator with a sinusoidally-corrugated slow-wave structure driven by a relativistic electron beam.}
\label{fig:RBWO_geom}
\end{figure}
\begin{figure*}
     \centering	
	\subfloat[\label{fig:beam}]{
     \includegraphics[width=5.5in]{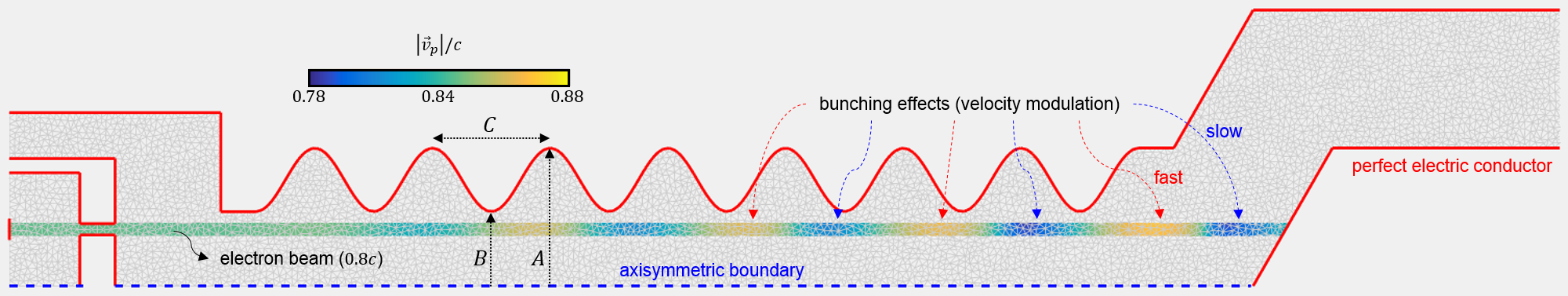}
	}
\\
	\subfloat[\label{fig:self_field}]{
	\includegraphics[width=5.5in]{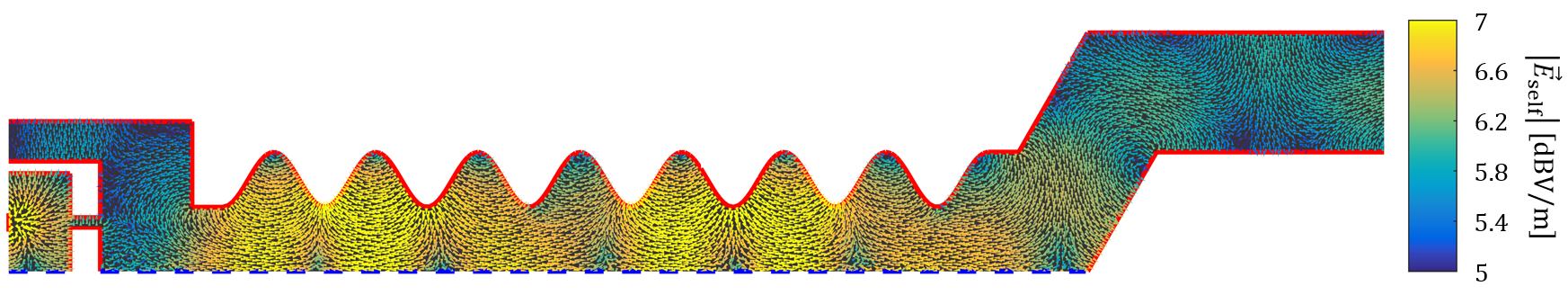}     
	}     
	\caption{Snapshots for (a) the velocity-modulated electron beam at $43.7$ ns and (b) the electric field (self-field) distribution at $83.3$ ns. The vertical axis is $\rho$ and horizontal axis is $z$.}
\label{fig:BWO}
\end{figure*}
\begin{figure}
    \centering
	\subfloat[\label{fig:ts}]{
    \includegraphics[width=2.25in]{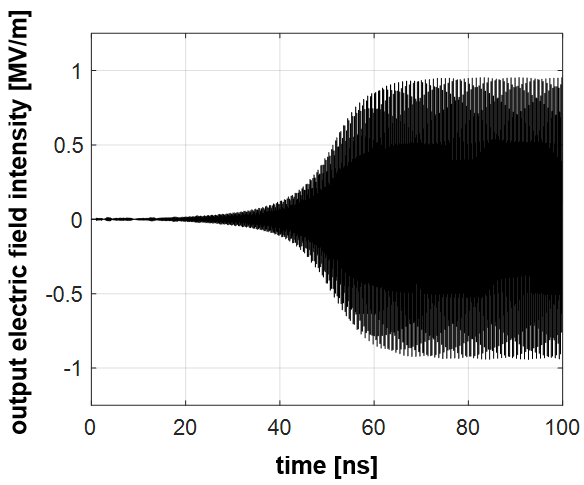}
    }
\\
	\subfloat[\label{fig:fs}]{
    \includegraphics[width=2.25in]{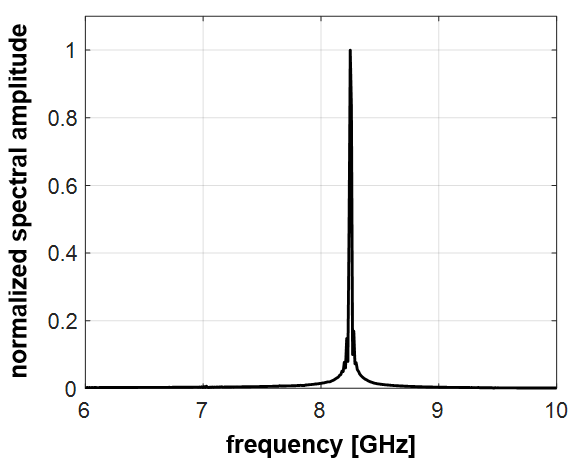}
    }
\caption{Output signals from the BWO device in (a) time domain and (b) frequency domain.}
\label{fig:output}
\end{figure}

In this section, we consider a backward-wave oscillator (BWO) driven by energetic electron beams in the relativistic regime designed to produce a high-power microwave signal \cite{gold1997review}, as depicted in Fig. \ref{fig:RBWO_geom}.
The proposed FETD-BOR solver is incorporated into a PIC algorithm~\cite{dawson1983particle,hockney1988computer,birdsall2004plasma} to simulate the wave-plasma interaction in the device~\cite{na2017axisymmetric}.
The PIC algorithm is based on an unstructured grid and explained in detail in~\cite{na2017axisymmetric,moon2015exact,na2016local}.
For this problem it suffices to consider the $\text{TE}^{\phi}$ polarized field with $m=0$.
In a relativistic BWO, the energy of space-charge modes is converted into microwaves via Cerenkov radiation~\cite{cairns1997generation}.
The BWO employs a slow-wave structure to produce such radiation~\cite{chipengo2015novel}.
In present case, the BWO system consists of a cathode, an anode, a slow-wave structure with sinusoidal corrugations, a beam collector, and a coaxial output port, as depicted in Fig.~\ref{fig:beam}.
The electron beam is produced by an external voltage between the cathode and anode.
In the slow-wave structure, the space charge modes evolve to $\text{TM}_{01}$ modes.
The oscillation of the modal field leads to beam velocity modulation and a quasi-periodic bunching of the electron beam distribution.
Lateral beam confinement is obtained by an externally applied static magnetic field. 
Coherent RF signals are detected and extracted at the coaxial RF output port as illustrated in Fig.~\ref{fig:beam}.
The outer radial boundary of the slow-wave structure is expressed as $R\left(z\right)=\left(A-B\right)\cos\left(2\pi z/C\right)+B$ where $A$ and $B$ are maximum and minimum radii, respectively, and $C$ is the axial corrugation period.
For X-band operation, we set $A=1.95$ cm, $B=1.05$ cm, and $C=1.67$ cm for a beam-velocity $v=2.5\times10^{8}$ m/s. 
The coaxial RF output port is truncated by a PML~\cite{donderici2008conformal,donderici2008mixed}.
The unstructured mesh has $N_{0}=2,892$, $N_{1}=8,155$, $N_{2}=5,264$, and $l_{\text{ave}}$=1.4468 mm where $l_{\text{ave}}$ is an average edge size.
The time step is $\Delta t=0.5$ ps {\color{black}corresponding to the Courant-Friedrichs-Lewy (CFL) number $0.5$}. 
As typical in PIC simulations, we employ a coarse-graining of the phase space, and each ``superparticle'' in the simulation represents $1.5\times10^{8}$ electrons. 
The resultant electron density $n_{e}$ yields a Debye length $\lambda_{D}=20.24$ mm.
The resulting number-density per Debye sphere $N_{D}$ equals $5.56\times 10^{11}$ particles and hence a collisionless plasma assumption is valid in this case. 
The self-field evolution and spectrum at the output port are shown in Figs.~\ref{fig:ts} and~\ref{fig:fs}, respectively.
From Fig.~\ref{fig:ts}, it is seen that the field grows to an RF oscillation near $50$ ns and saturates at around $100$ ns. The output signal has a peak at $8.27$ GHz, as shown in Fig.~\ref{fig:fs}.
Fig. \ref{fig:beam} shows the electron beam distribution at $43.7$ ns. The velocity of each particle is color-encoded and the bunching effect due to velocity modulation is clearly visible.
Fig. \ref{fig:self_field} also illustrates the steady-state profile of the BWO system at $83.3$ ns. The vector plot of the corresponding self-fields clearly shows that a strong $\text{TM}_{01}$ mode is indeed present.
}

{\color{black}
\section{Conclusion}
We presented a new finite-element time-domain (FETD) Maxwell solver for the analysis of body-of-revolution (BOR) geometries. The proposed solver is based on discrete exterior calculus (DEC) and transformation optics (TO) concepts.
We explored TO principles to map the original 3-D problem from a cylindrical coordinate system to an equivalent problem on a 2-D (Cartesian-like) meridian $\rho z$ plane, where the cylindrical metric is factored out from the differential operators and embedded on an effective (artificial) inhomogeneous and anisotropic medium that fills the domain.
This enables the use of Cartesian 2-D FE code with no modifications necessary except to accommodate the presence of anisotropic media. 
The spatial discretization is done on an unstructured mesh on the 2-D meridian plane and effected by 
decomposing the fields into their  $\text{TE}^{\phi}$ and $\text{TM}^{\phi}$ components and expanding each eigenmode into an appropriate set of (vector or scalar) basis functions (Whitney forms) based on DEC principles. A leap-frog  (symplectic) time-integrator is applied to the semi-discrete Maxwell curl equations and used to obtain a fully discrete, marching-on-time evolution algorithm.
Unlike prior solvers,  the present FETD-BOR Maxwell solver does not require any modifications on the basis functions adjacent to the symmetry axis. Rather, the field behavior on the symmetry axis can be simply implemented through properly selected homogeneous Dirichlet and Neumann applied to the eigenmodal expansion.

\section*{Acknowledgments}
This work was supported in part by National Science Foundation grant ECCS-1305838, Department of Defense, Defense Threat Reduction Agency grant HDTRA1-18-1-0050, Ohio Supercomputer Center grants PAS-0061 and PAS-0110, 
S\~{a}o Paulo State Research Foundation (FAPESP) grant 2015/50268-5, and the Ohio State University Presidential Fellowship program.

The content of the information does not necessarily reflect the position or the policy of the U.S. federal government, and no official endorsement should be inferred.
}

{\color{black}
\clearpage
\appendix

\section{Whitney forms and pairing operations}
\label{ap:Forms}
 Whitney $p$-forms are canonical interpolants of discrete differential $p$-forms~\cite{whitneybook}.
As explained below, 
 Whitney $p$-forms are naturally paired to the $p$-cells of the mesh, where $p$ refers to the dimensionality, i.e. $p=0$ refers to nodes, $p=1$ to edges, $p=1$ to facets and so on~\cite{teixeira1999lattice}. On simplices (e.g. on triangular cells in 2-D or tetrahedral cells in 3-D), Whitney 
 0-, 1-, and 2-forms are expressed as~\cite{teixeira1999lattice, bossavit1988whitney,whitneybook}
\begin{flalign}
{w}_{i}^{(0)}&=
\lambda_{i},
\\
w_{i}^{(1)}&=
\lambda_{i_{a}} d \lambda_{i_{b}}-\lambda_{i_{b}} d \lambda_{i_{a}},
\\
w_{i}^{(2)}&=
2
\left(
\lambda_{i_{a}} d \lambda_{i_{b}}\wedge d \lambda_{i_{c}}+
\lambda_{i_{b}} d \lambda_{i_{c}}\wedge d \lambda_{i_{a}}
+\lambda_{i_{c}} d \lambda_{i_{a}}\wedge d  \lambda_{i_{b}}\right),
\end{flalign}
where $d$ is the exterior derivative, $\wedge$ is the exterior product, $i_{a}$, $i_{b}$, and $i_{c}$ denote the grid nodes belonging to the $i$-th $p$-cell for $p=1$ or $2$, and $\lambda$ denotes the barycentric coordinate associated to a given node. 

The corresponding vector proxies for Whitney 0-, 1-, and 2-forms write as~\cite{moon2015exact,teixeira1999lattice}
\begin{flalign}
{\text{W}}_{i}^{(0)}&=
\lambda_{i},
\\
\mathbf{W}_{i}^{(1)}&=
\lambda_{i_{a}}\mathbf{\nabla}\lambda_{i_{b}}-\lambda_{i_{b}}\mathbf{\nabla} \lambda_{i_{a}},
\\
\mathbf{W}_{i}^{(2)}&=
2
\left(
\lambda_{i_{a}}\mathbf{\nabla}\lambda_{i_{b}}\times\mathbf{\nabla} \lambda_{i_{c}}+
\lambda_{i_{b}}\mathbf{\nabla}\lambda_{i_{c}}\times\mathbf{\nabla} \lambda_{i_{a}}
+\lambda_{i_{c}}\mathbf{\nabla}\lambda_{i_{a}}\times\mathbf{\nabla} \lambda_{i_{b}}\right),
\end{flalign}

One of the key properties of Whitney $p$-forms is that they admit a natural ``pairing'' with the $p$-cells of the mesh~\cite{teixeira1999lattice}. 
Computationally, the pairing operation between an $i$-th $p$-cell of the grid $\sigma_{(p)}^{i}$ and 
a Whitney form $w^{(p)}_{j}$ associated with the $j$-th $p$-cell is effected by the integral below and yields~\cite{teixeira1999lattice,teixeira2014lattice}
\begin{eqnarray}
\left<\sigma_{(p)}^{i},w_{j}^{(p)}\right>=\int_{\sigma_{(p)}^{i}} w^{(p)}_{j}=\delta_{i,j},
\label{eq:paring}
\end{eqnarray}
where $\delta_{i,j}$ is the Kronecker delta, for $p=0, \ldots, 3$ in 3-D space.

\section{Generalized Stokes' theorem}
\label{ap:DEC1}
The generalized Stokes' theorem of exterior calculus~\cite{teixeira1999lattice,teixeira2014lattice,Kettunen2007,cairns1936the,Kettunen1999b} states
\begin{eqnarray}
\left<\sigma_{(p+1)},d w^{(p)}_{j}\right>
=
\left<\left(\partial \sigma_{(p+1)}\right)_{(p)},w^{(p)}_{j}\right>
\label{eq:GST}
\end{eqnarray}
where $\partial$ is the boundary operator that maps an (oriented) $p$-cell on the grid to the set of (oriented) $(p-1)$-cells comprising its boundary. Note that 
 $\partial^2=0$ and hence $d^2=0$ from (\ref{eq:GST}). This latter identity is the exterior calculus counterpart of the
vector calculus identities $\mathbf{\nabla} \times \mathbf{\nabla} = \mathbf{0}$  and $\mathbf{\nabla} \cdot \mathbf{\nabla} \times = 0$.

The generalized Stokes' theorem recovers  Stokes' and Gauss' theorems of vector calculus for $p=1,2$, respectively, and the fundamental theorem of calculus for $p=0$.

\section{Discrete Maxwell's equations}
\label{ap:DEC2}
By pairing Faraday's law for the $\text{TE}^{\phi}$ field set in (\ref{eq:diff_FL_TE_pol}) with $K$-th $2$-cells $\sigma_{(2)}^{K}$ of the FE grid (primal mesh)
and applying the generalized Stokes' theorem, we obtain
\begin{flalign}
&\left<\sigma^{K}_{(2)},\sum_{j=1}^{N_{1}}
\mathbb{E}_{j,m}^{\parallel}\left(t\right) 
\left[{d'}^{\parallel}w_{j}^{(1)}\right]\right>
=
-\left<\sigma^{K}_{(2)},\frac{\partial}{\partial t}
\sum_{k=1}^{N_{2}}
\mathbb{B}_{k,m}^{\perp}\left(t\right) w_{k}^{(2)}\right>,
\label{eq:paring_GST_FL_1}
\end{flalign}
\begin{flalign}
&
\left<
\left(\partial \sigma^{K}_{(2)}\right)_{(1)},\sum_{j=1}^{N_{1}}
\mathbb{E}_{j,m}^{\parallel}\left(t\right) 
w_{j}^{(1)}\right>
=
-\left<\sigma^{K}_{(2)},\frac{\partial}{\partial t}
\sum_{k=1}^{N_{2}}
\mathbb{B}_{k,m}^{\perp}\left(t\right) w_{k}^{(2)}\right>.
\label{eq:paring_GST_FL_2}
\end{flalign}
Using $\left(\partial \sigma^{K}_{(2)}\right)_{(1)}=\sum_{j=1}^{N_{1}}C_{K,j}\sigma^{j}_{(1)}$, where $C_{K,j}$ is the incidence matrix associated to the exterior derivative applied to 1-forms (curl operator on the mesh), see  \ref{ap:incidence}), we obtain~\cite{teixeira1999lattice,teixeira2014lattice,hughes1981lagrangian,guth1980existence}
\begin{flalign}
\sum_{j=1}^{N_{1}} C_{K,j} \mathbb{E}_{j,m}^{\parallel}\left(t\right) =-\frac{\partial }{\partial t} \mathbb{B}_{K,m}^{\perp}\left(t\right) ,
\label{eq:paring_GST_FL_3}
\end{flalign}
for $m=-M_{\phi},...,M_{\phi}$.
The elements of the incidence matrix take values  in the set of $\left\{-1,0,1\right\}$,

Likewise, pairing (\ref{eq:diff_FL_TM_pol}) with $J$-th $1$-cells $\sigma_{(1)}^{J}$ of the primal mesh gives
\begin{flalign}
& \!\!\!\!\!\!\!\!\!\!
{\color{black}
\left<
\sigma^{J}_{(1)},\sum_{i=1}^{N_{0}} \mathbb{E}_{i,m}^{\perp}\left(t\right) 
\left[{d'}^{\parallel}w_{i}^{(0)}\right]
\right>
-
\left<\sigma^{J}_{(1)},\left|m\right|\sum_{j=1}^{N_{1}} \mathbb{E}_{j,m}^{\parallel}\left(t\right) 
w_{j}^{(1)}
\right>
}
\nonumber \\
&~~~~~~~~~~~~~~~~~~~~~~~~~~~~~~~~~~~~~~~~~~~
=
-\left<\sigma^{J}_{(1)},\frac{\partial}{\partial t}
\sum_{j=1}^{N_{1}} \mathbb{B}_{j,m}^{\parallel}\left(t\right) w_{j}^{(1)}\right>,
\label{eq:paring_GST_FL_1}
\end{flalign}
and applying generalized Stokes' theorem to the left-hand side of (\ref{eq:paring_GST_FL_1}) yields 
\begin{flalign}
&\!\!\!\!
{\color{black}
\left<\left(\partial\sigma^{J}_{(1)}\right)_{(0)},\sum_{i=1}^{N_{0}}
\mathbb{E}_{i,m}^{\perp}\left(t\right) 
w_{i}^{(0)}\right>
-
\left<\sigma^{J}_{(1)},\left|m\right|\sum_{j=1}^{N_{1}}
\mathbb{E}_{j,m}^{\parallel}\left(t\right) w_{j}^{(1)}\right>
}
\nonumber \\
&~~~~~~~~~~~~~~~~~~~~~~~~~~~~~~~~~~~~~~~~~~~~~~~~=
-\left<\sigma^{J}_{(1)},\frac{\partial}{\partial t}
\sum_{j=1}^{N_{1}}
\mathbb{B}_{j,m}^{\parallel}\left(t\right) w_{j}^{(1)}\right>,
\label{eq:paring_GST_FL_2}
\end{flalign}
Similarly to before, we can write $\left(\partial \sigma^{J}_{(1)}\right)_{(0)}=\sum_{i=1}^{N_{0}}G_{J,i}\sigma^{i}_{(1)}$, where $G_{J,i}$ is the incidence matrix associated to the exterior derivative applied to 0-forms (gradient operator on the mesh), and obtain
\begin{flalign}
{\color{black}
\sum_{i=1}^{N_{0}} G_{J,i} \mathbb{E}_{i,m}^{\perp}\left(t\right)-\left|m\right|\mathbb{E}_{J,m}^{\parallel}\left(t\right) =-\frac{\partial}{\partial t} \mathbb{B}_{J,m}^{\parallel}\left(t\right) ,
\label{eq:paring_GST_FL_3}
}
\end{flalign}
An analogous procedure can be used to obtain the discrete rendering of Ampere's law for on the dual mesh.

{\color{black}
\section{Incidence Matrices} 
\label{ap:incidence}
\begin{figure}[t]
\centering
\includegraphics[width=2.75in]{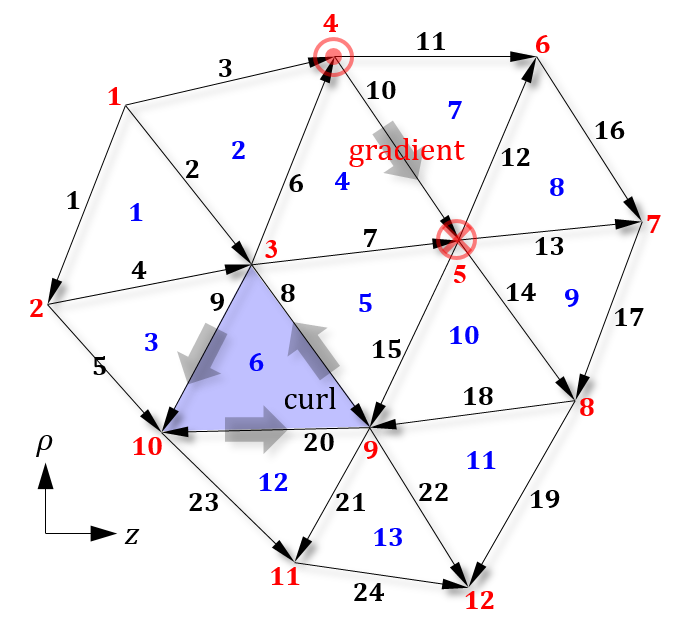}
\caption{Example (primal) unstructured mesh.}
\label{fig:mesh_example}
\end{figure}
\begin{figure}[t]
\centering
\subfloat[\label{fig:inc_mat}]{	\includegraphics[width=5in]{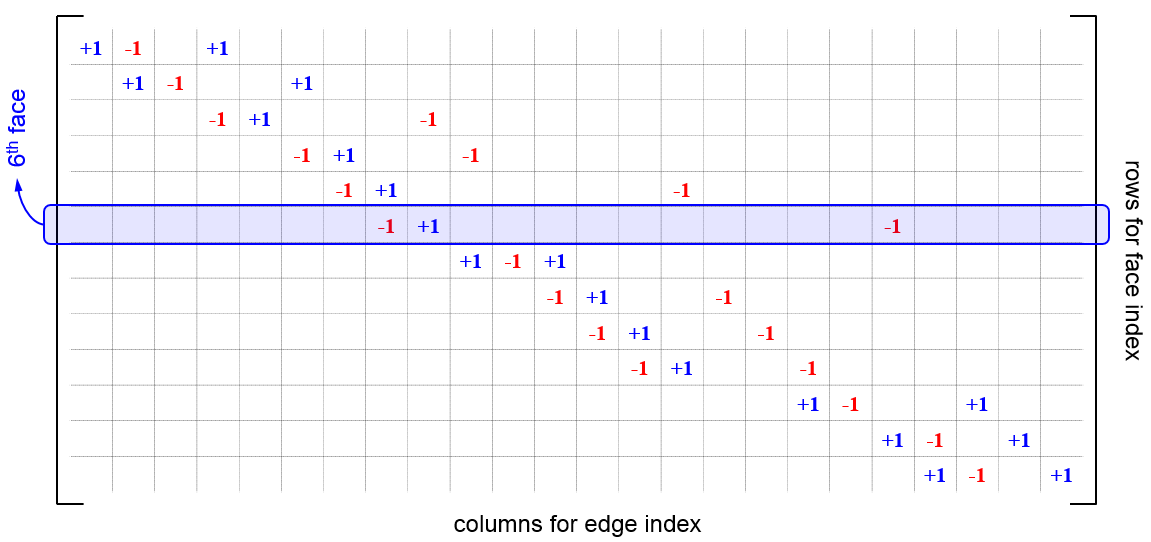}}
\\
\subfloat[\label{fig:grad_mat}]{	\includegraphics[width=2.95in]{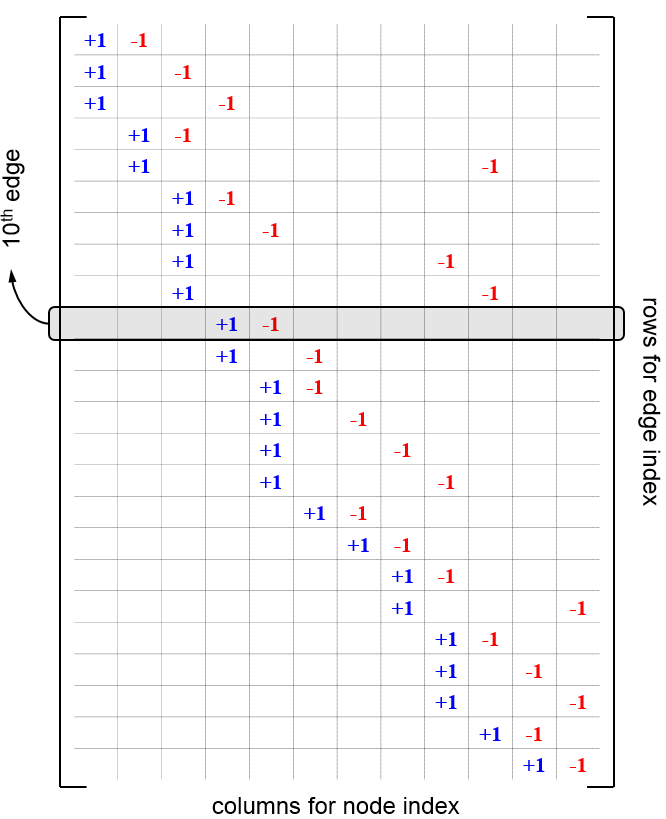}}
\caption{Incidence matrices for (a) curl $\left[\mathcal{D}_{\text{curl}}\right]$ and (b) gradient $\left[{\mathcal{D}}_{\text{grad}}\right]$ operators for the mesh in Fig. \ref{fig:mesh_example}.}
\label{fig:incidence_matrices}
\end{figure}
Incidence matrices can be used to represent on a mesh the discrete exterior derivative or, equivalently, the grad, curl, and div operators distilled from their metric structure~\cite{teixeira1999lattice,teixeira2014lattice,Kettunen1999b}. 
Since, from~(\ref{eq:GST}), the discrete exterior derivative can be seen as the dual of the boundary operator, incidence matrices encode the relationship between each oriented $p$-cell of the mesh and its boundary oriented $(p-1)$-cells (say, between an edge and its boundary nodes, a face element and its boundary edges, and so on). 
To provide a concrete example, we consider a small mesh with perfect magnetic conductor (or free edges) boundaries as depicted in Fig. \ref{fig:mesh_example}.
Red-colored numbers denote the nodal indices, black-colored numbers the edge indices, and blue-colored numbers the face indices. 
Intrinsic edge orientation is defined by ascending index order of the two nodes associated with any given edge.
For example, if we consider $\left[\mathcal{D}_{\text{curl}}\right]$, of size $N_{2} \times N_{1}$, there are three edges wrapping face number 6: edges 8, 9, and 20. 
{\color{black}
As a result, $\left[\mathcal{D}_{\text{curl}}\right]_{6,8}=1$, $\left[\mathcal{D}_{\text{curl}}\right]_{6,9}=-1$, and $\left[\mathcal{D}_{\text{curl}}\right]_{6,9}=1$. 
}
The sign is determined by comparing the intrinsic orientation of each edge with the curl in Fig. \ref{fig:mesh_example}: if they are opposite, the element is $-1$, otherwise it is $+1$. 
Furthermore, $\left[\mathcal{D}_{\text{curl}}\right]_{6,j}=0$ for all other $j-$th edges. This is represented in Fig. \ref{fig:inc_mat}, which shows the entire  $\left[\mathcal{D}_{\text{curl}}\right]$ for this mesh.
{\color{black}A curl orientation on each face is supposed to follow the intrinsic orientation of the first local edge (i.e. an edge with the smallest index among three edges for the face).}
Likewise, if we consider $\left[{\mathcal{D}}_{\text{grad}}\right]$, of size $N_{1} \times N_{0}$, there are two nodes connected to edge 10: nodes 4 and 5. 
{\color{black}
The corresponding elements are $\left[\mathcal{D}_{\text{grad}}\right]_{10,4}=-1$ and $\left[\mathcal{D}_{\text{grad}}\right]_{10,5}=1$. The element for the diverging node with the gradient (the intrinsic edge orientation) in Fig. \ref{fig:mesh_example} is $-1$, otherwise it is $+1$.
}
}

{\color{black}
\section{Discrete Hodge Matrix} \label{ap:Hodge}
A (discrete) Hodge star operator encodes all metric information and is used to transfer information between the primal and dual meshes~\cite{teixeira1999lattice,Kettunen2007,he2006geometric,Gillette2011,Kettunen1999}. 
Here, we use a Galerkin-Hodge construction~\cite{Kettunen2007,kim2011parallel,Dodziuk,Kettunen1999}, which leads to symmetric positive definite matrices and enables energy-conserving discretizations with standard local energy positivity {\color{black}in arbitrary simplicial meshes~\cite{teixeira2014lattice}. As noted in Section 2, the Galerkin-Hodge operator is {\it not} a natural consequence of DEC~\cite{Kotiuga1}.}

The Hodge operator also incorporates the constitutive properties (permittivity and permeability) of the background medium~\cite{donderici2008mixed}.
Inhomogeneous and anisotropic media can be easily dealt with by incorporating piecewise constant permittivity and permeability over each cell, for example.
In the present FETD-BOR solver, the elements of the Hodge matrices including the radial scaling factor from the cylindrical metric are assembled by adding the contributions from all cells as:
\begin{flalign}
&\left[\star_{\epsilon}\right]_{J,j}^{1\rightarrow1}=
\sum_{k=1}^{N_{2}}
\int_{\Omega_{k}} \left({\epsilon_k \rho_{k} }\right)\mathbf{W}_{J}^{(1)} \cdot \mathbf{W}_{j}^{(1)} dV,
\\
&\left[\star_{\mu^{-1}}\right]_{K,k}^{2\rightarrow2}=
\sum_{k=1}^{N_{2}}
\int_{\Omega_{k}} \left({\mu^{-1}_k \rho_{k} }\right)\mathbf{W}_{K}^{(2)} \cdot \mathbf{W}_{k}^{(2)} dV,
\\
&\left[\star_{\epsilon}\right]_{I,i}^{0\rightarrow0}=
\sum_{k=1}^{N_{2}}
\int_{\Omega_{k}}\left({\epsilon_k \rho^{-1}_{k}}\right)\left[\text{W}_{I}^{(0)}\hat{\phi}\right] \cdot \left[\text{W}_{i}^{(0)}\hat{\phi}\right] dV,
\\
&
{\color{black}
\left[\star_{\mu^{-1}}\right]_{J,j}^{1\rightarrow1}=
\sum_{k=1}^{N_{2}}
\int_{\Omega_{k}}\left(\mu_k^{-1} \rho^{-1}_{k}\right)\left[\mathbf{W}_{J}^{(1)}\times\hat{\phi}\right] \cdot \left[\mathbf{W}_{j}^{(1)}\times\hat{\phi}\right] dV,
}
\end{flalign}
where $\Omega_{k}$ is the area of the $k-$th cell, and $\rho_{k}=\sum_{i=1}^{3}\rho_{k_{i}}/3$ where $\rho_{k_{i}}$ is $\rho$ coordinate of $i-$th node touching $k-$th face and for simplicity we have assumed isotropic media assuming permittivity and permeability values $\epsilon_k$ and $\mu_k$, resp., on cell $k$.
Since Whitney forms have compact support, we can express the global discrete Hodge matrix as a sum of local matrices (excluding element-wise permittivity and permeability information) for the $K$-th face as
\begin{eqnarray}
&&\!\!\!\!\!\!\!\!\!\!\!\!\!\!\!\!\!\!\!\!\!\!\!\!
\left[\mathcal{T}\right]_{K}^{0\rightarrow0}=\Delta_{K}
\begin{bmatrix}
1/6 & 1/12 & 1/12 \\ 
1/12 & 1/6 & 1/12 \\ 
1/12 & 1/12 & 1/6
\end{bmatrix},
\\
&&\!\!\!\!\!\!\!\!\!\!\!\!\!\!\!\!\!\!\!\!\!\!\!\!
\left[\mathcal{T}\right]_{K}^{1\rightarrow1}=\Delta_{K}
\begin{bmatrix}
T_{11}^{1\rightarrow1} & T_{12}^{1\rightarrow1} & T_{13}^{1\rightarrow1} \\ T_{21}^{1\rightarrow1} & T_{22}^{1\rightarrow1} & T_{23}^{1\rightarrow1} \\ T_{31}^{1\rightarrow1} & T_{32}^{1\rightarrow1} & T_{33}^{1\rightarrow1}
\end{bmatrix},
\\
&&\!\!\!\!\!\!\!\!\!\!\!\!\!\!\!\!\!\!\!\!\!\!\!\!
\left[\mathcal{T}\right]_{K}^{2\rightarrow2}=4\Delta_{K}\left(\mathbf{\nabla}\lambda_{1}\times\mathbf{\nabla}\lambda_{2}\right)\cdot\hat{\phi},
\end{eqnarray}
where $\Delta_{K}$ is the area of $K$-th face and 
\begin{flalign}
&T_{11}^{1\rightarrow1}
= 	\frac{\mathbf{\nabla}\lambda_{1} \cdot \mathbf{\nabla}\lambda_{1}}{6} 
+ 	\frac{\mathbf{\nabla}\lambda_{2} \cdot \mathbf{\nabla}\lambda_{2}}{6} 
- 	\frac{\mathbf{\nabla}\lambda_{1} \cdot \mathbf{\nabla}\lambda_{2}}{6},  \\
&T_{12}^{1\rightarrow1}
= 	\frac{\mathbf{\nabla}\lambda_{1} \cdot \mathbf{\nabla}\lambda_{1}}{6} 
- 	\frac{\mathbf{\nabla}\lambda_{2} \cdot \mathbf{\nabla}\lambda_{2}}{6} 
- 	\frac{\mathbf{\nabla}\lambda_{1} \cdot \mathbf{\nabla}\lambda_{2}}{6},  \\
&T_{13}^{1\rightarrow1}
= 	\frac{\mathbf{\nabla}\lambda_{1} \cdot \mathbf{\nabla}\lambda_{1}}{6} 
- 	\frac{\mathbf{\nabla}\lambda_{2} \cdot \mathbf{\nabla}\lambda_{2}}{6} 
+ 	\frac{\mathbf{\nabla}\lambda_{1} \cdot \mathbf{\nabla}\lambda_{2}}{6},  \\
&T_{21}^{1\rightarrow1}=T_{12}^{1\rightarrow1},\\
&T_{22}^{1\rightarrow1}
= 	\frac{\mathbf{\nabla}\lambda_{1} \cdot \mathbf{\nabla}\lambda_{1}}{2} 
+ 	\frac{\mathbf{\nabla}\lambda_{2} \cdot \mathbf{\nabla}\lambda_{2}}{6} 
+ 	\frac{\mathbf{\nabla}\lambda_{1} \cdot \mathbf{\nabla}\lambda_{2}}{2},  \\
&T_{23}^{1\rightarrow1}
= 	\frac{\mathbf{\nabla}\lambda_{1} \cdot \mathbf{\nabla}\lambda_{1}}{6} 
+ 	\frac{\mathbf{\nabla}\lambda_{2} \cdot \mathbf{\nabla}\lambda_{2}}{6} 
+ 	\frac{\mathbf{\nabla}\lambda_{1} \cdot \mathbf{\nabla}\lambda_{2}}{2},  \\
&T_{31}^{1\rightarrow1}=T_{13}^{1\rightarrow1}, \\
&T_{32}^{1\rightarrow1}=T_{23}^{1\rightarrow1}, \\
&T_{33}^{1\rightarrow1}
= 	\frac{\mathbf{\nabla}\lambda_{1} \cdot \mathbf{\nabla}\lambda_{1}}{6} 
+ 	\frac{\mathbf{\nabla}\lambda_{2} \cdot \mathbf{\nabla}\lambda_{2}}{2} 
+ 	\frac{\mathbf{\nabla}\lambda_{1} \cdot \mathbf{\nabla}\lambda_{2}}{2}.
\end{flalign}
Due to the local support of  the Whitney forms, the above Hodge matrices are very sparse (and diagonally dominant).
Their sparsity patterns for the mesh in Fig. \ref{fig:mesh_example} are provided in Fig. \ref{fig:sp_DHM}. 
The number of non-zero elements per row (or column) in these Hodge matrices is invariant with respect to the mesh size, so the sparsity increases for larger meshes.
\begin{figure}
\centering
\subfloat[\label{fig:eps_mat}]{\includegraphics[width=2in]{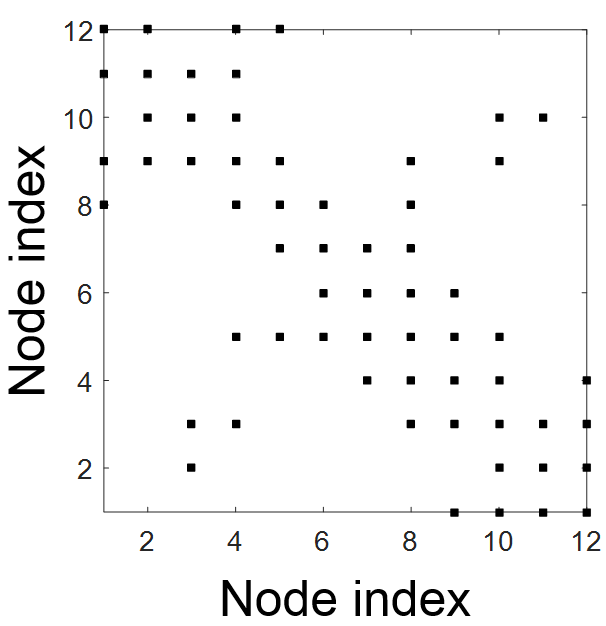}}
~
\subfloat[\label{fig:eps_mat}]{\includegraphics[width=2in]{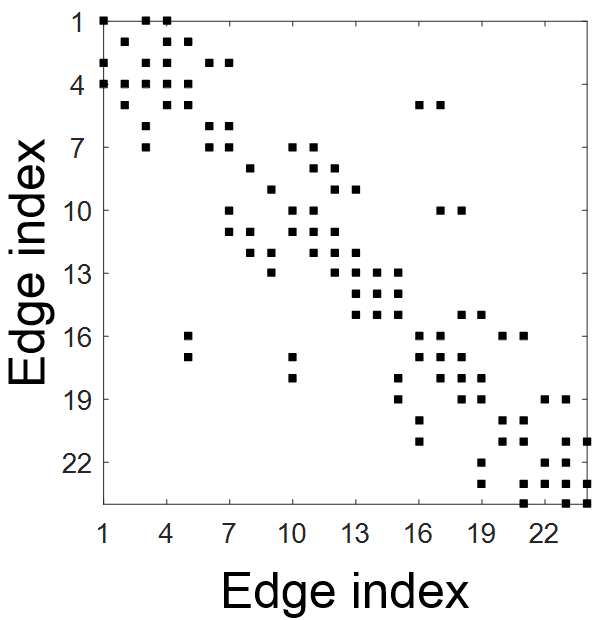}}
\\
\subfloat[\label{fig:eps_mat}]{\includegraphics[width=2in]{NF/eps_mat.PNG}}
~
\subfloat[\label{fig:mu_mat}]{\includegraphics[width=2in]{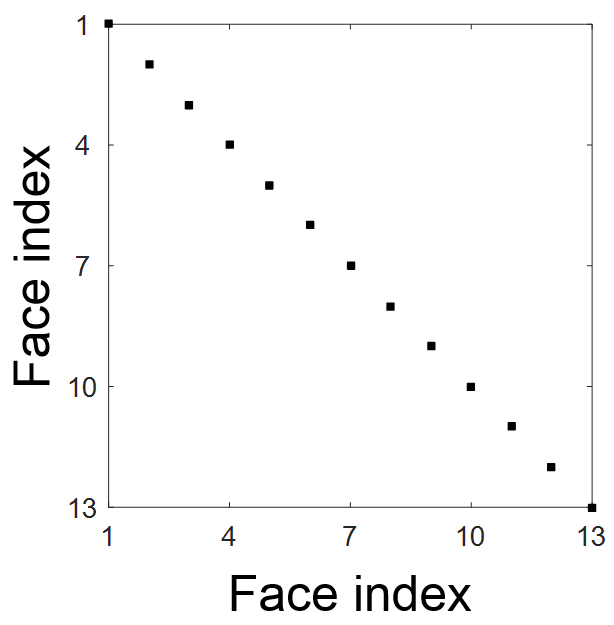}}
\caption{Sparsity patterns for discrete Hodge matrices corresponding to the toy mesh depicted in Fig. \ref{fig:mesh_example}:
(a) $\left[\star_{\epsilon}\right]^{0\rightarrow0}$, (b) $\left[\star_{\epsilon}\right]^{1\rightarrow1}$, (c) $\left[\star_{\mu}^{-1}\right]^{1\rightarrow1}$, and (d) $\left[\star_{\mu^{-1}}\right]^{2\rightarrow2}$.}
\label{fig:sp_DHM}
\end{figure}
}

{\color{black}
\section{Cartesian-like PML implementation} \label{ap:PML}
A perfectly matched layer (PML) is used to absorb outgoing waves in FE simulations, enabling analysis of open-domain problems~\cite{Berenger,ComplexPML}.
As described before, in the present FETD-BOR the spatial discretization is performed in the meridian plane mapped onto a Cartesian domain with the cylindrical metric factor transferred to the constitutive relations. The resulting constitutive relations correspond to a medium that is inhomogeneous and doubly anisotropic. As such, a Cartesian PML implementation extended to such media can be used. Such formulation exists~\cite{teixeira1998general} and is adapted here to the FETD-BOR case as follows.

In the 2-D Cartesian plane, the PML can be effected as an analytic continuation on the spatial variables to complex space~\cite{ComplexPML,teixeira1998general}, given by
 $u\rightarrow\tilde{u}=\int_{0}^{u}s_{u}\left(u'\right)du'$ where $s_{u}\left(u'\right)$ is a complex stretching variable
and $u$ stands for $\rho$ or $z$.
This transformation can also be expressed as
\begin{flalign}
{{\mathbf{r}}}^{' \parallel}\rightarrow{\tilde{\mathbf{r}}}^{' \parallel}&=\bar{\bar{\mathbf{\Gamma}}}\cdot\mathbf{r}^{' \parallel},
\label{eq:ap_transform}
\end{flalign}
where $\bar{\bar{\mathbf{\Gamma}}}=\hat{\rho}\hat{\rho}\left({\tilde{\rho}}/{\rho}\right)+\hat{z}\hat{z}\left({\tilde{z}}/{z}\right)$.
As before, the apostrophe $'$ in ${{\mathbf{r}}}^{' \parallel}$ denotes the transverse coordinates on the 2-D meridian plane.
The modified nabla operator (posterior to the TO-based transformation and hence devoid of the $1/\rho$ factor in the $\phi$ derivative) following such analytical continuation is given by
\begin{flalign}
{\nabla}'\rightarrow\tilde{\nabla}'=\hat{\rho}\frac{1}{s_{\rho}}\frac{\partial}{\partial \rho}+\hat{\phi}\frac{\partial}{\partial \phi}+\hat{z}\frac{1}{s_{z}}\frac{\partial}{\partial z},
\end{flalign}
or simply 
\begin{flalign}
\tilde{\nabla}'=\bar{\bar{\mathbf{S}}}\cdot{\nabla}', 
\label{eq:ap_analytic_cont_simp}
\end{flalign}
where $\bar{\bar{\mathbf{S}}}=\hat{\rho}\hat{\rho}\left(1/s_{\rho}\right)+\hat{\phi}\hat{\phi}\left(1\right)+\hat{z}\hat{z}\left(1/s_{z}\right)$.
Following~\cite{teixeira1998general}, since $s_{u}\left(u\right)$ and $\partial/\partial u'$ commute when $u\neq u'$ and $\bar{\bar{\mathbf{S}}}$ is a diagonal tensor, the following identity holds for any vector $\mathbf{a}$ in the Cartesian-like 2-D meridian plane:
\begin{flalign}
{\nabla'}\times\left(\bar{\bar{\mathbf{S}}}^{-1}\cdot\mathbf{a}\right)
=
\left(\text{det}\bar{\bar{\mathbf{S}}}\right)^{-1}\bar{\bar{\mathbf{S}}}\cdot\left(\bar{\bar{\mathbf{S}}}\cdot{\nabla'}\right)\times\mathbf{a}.
\label{eq:ap_iden_PML}
\end{flalign}

Applying this analytic continuation to (\ref{eq:FL_eqv_TE_pol}), (\ref{eq:FL_eqv_TM_pol}), (\ref{eq:AL_TM_pol_vec}), and (\ref{eq:AL_TE_pol_vec}) in the Fourier domain (with time convention of $e^{j\omega t}$) yields the modified Maxwell's equations for each mode $m$ as
\begin{flalign}
\tilde{\nabla}'^{\parallel}\times \mathbf{E'}^{\parallel c}_{m}\left({\tilde{\mathbf{r}}}^{'\parallel}\right)&=-j\omega\mathbf{B'}^{\perp c}_{m}\left({\tilde{\mathbf{r}}}^{'\parallel}\right),
\label{eq:ap_NME_FL_TE_pol}
\\
{\color{black}
\tilde{\nabla}'^{\parallel}\times \mathbf{E'}^{\perp c}_{m}\left({\tilde{\mathbf{r}}}^{'\parallel}\right)
}
&
{\color{black}
=-j\omega\mathbf{B'}^{\parallel c}_{m}\left({\tilde{\mathbf{r}}}^{'\parallel}\right) + \left|m\right|\mathbf{E'}^{\parallel c}_{m}\left({\tilde{\mathbf{r}}}^{'\parallel}\right)\times\hat{\phi},
\label{eq:ap_NME_FL_TM_pol}
}
\\
\tilde{\nabla}'^{\parallel}\times \mathbf{H'}^{\parallel c}_{m}\left({\tilde{\mathbf{r}}}^{'\parallel}\right)&=j\omega\mathbf{D'}^{\perp c}_{m}\left({\tilde{\mathbf{r}}}^{'\parallel}\right),
\label{eq:ap_NME_AL_TM_pol}
\\
{\color{black}
\tilde{\nabla}'^{\parallel}\times \mathbf{H'}^{\perp c}_{m}\left({\tilde{\mathbf{r}}}^{'\parallel}\right)
}
&
{\color{black}
=j\omega \mathbf{D'}^{\parallel c}_{m}\left({\tilde{\mathbf{r}}}^{'\parallel}\right) - \left|m\right|\mathbf{H'}^{\parallel c}_{m}\left({\tilde{\mathbf{r}}}^{'\parallel}\right)\times\hat{\phi},
}
\label{eq:ap_NME_AL_TE_pol}
\end{flalign}
with constitutive relations in analytic-continued complex space as
\begin{flalign}
\mathbf{D'}^{c}_{m}\left({\tilde{\mathbf{r}}}^{'\parallel}\right)&=\bar{\bar{\epsilon}}'\left(\omega\right)\cdot\mathbf{E'}^{c}_{m}\left({\tilde{\mathbf{r}}}^{'\parallel}\right),
\\
\mathbf{B'}^{c}_{m}\left({\tilde{\mathbf{r}}}^{'\parallel}\right)&=\bar{\bar{\mu}}'\left(\omega\right)\cdot\mathbf{H'}^{c}_{m}\left({\tilde{\mathbf{r}}}^{'\parallel}\right),
\end{flalign}
where the superscript $c$ denotes non-Maxwellian (complex space) fields and $\bar{\bar{\epsilon}}'$ and $\bar{\bar{\mu}}'$ indicates constitutive parameters of the original medium incorporating the radial scaling factors from the TO mapping.
Next, using (\ref{eq:ap_transform}) and (\ref{eq:ap_analytic_cont_simp}), we can revert (\ref{eq:ap_NME_FL_TE_pol})$-$(\ref{eq:ap_NME_AL_TE_pol}) back to a real-valued spatial domain by writing
\begin{flalign}
&\!\!\!\!\!\!
\left(\bar{\bar{\mathbf{S}}}\cdot{\nabla}'^{\parallel}\right)
\times
\mathbf{E'}^{\parallel c}_{m}  \left(\bar{\bar{\mathbf{\Gamma}}}\cdot{{\mathbf{r}}}^{'\parallel}\right)
=
-j\omega\mathbf{B'}^{\perp c}_{m}  \left(\bar{\bar{\mathbf{\Gamma}}}\cdot{{\mathbf{r}}}^{'\parallel}\right),
\label{eq:ap_NME_FL_mod_TE_pol}
\\
&\!\!\!\!\!\!
\left(\bar{\bar{\mathbf{S}}}\cdot{\nabla}'^{\parallel}\right)
\times \mathbf{E'}^{\perp c}_{m}  \left(\bar{\bar{\mathbf{\Gamma}}}\cdot{{\mathbf{r}}}^{'\parallel}\right)
=
-j\omega\mathbf{B'}^{\parallel c}_{m}  \left(\bar{\bar{\mathbf{\Gamma}}}\cdot{{\mathbf{r}}}^{'\parallel}\right)
-
\left|m\right|\hat{\phi}\times\mathbf{E'}^{\parallel c}_{m}  \left(\bar{\bar{\mathbf{\Gamma}}}\cdot{{\mathbf{r}}}^{'\parallel}\right),
\label{eq:ap_NME_FL_mod_TM_pol}
\\
&\!\!\!\!\!\!
\left(\bar{\bar{\mathbf{S}}}\cdot{\nabla}'^{\parallel}\right)
\times
\mathbf{H'}^{\parallel c}_{m}  \left(\bar{\bar{\mathbf{\Gamma}}}\cdot{{\mathbf{r}}}^{'\parallel}\right)
=
j\omega\mathbf{D'}^{\perp c}_{m}  \left(\bar{\bar{\mathbf{\Gamma}}}\cdot{{\mathbf{r}}}^{'\parallel}\right),
\label{eq:ap_NME_AL_mod_TM_pol}
\\
&\!\!\!\!\!\!
\left(\bar{\bar{\mathbf{S}}}\cdot{\nabla}'^{\parallel}\right)
\times\mathbf{H'}^{\perp c}_{m}  \left(\bar{\bar{\mathbf{\Gamma}}}\cdot{{\mathbf{r}}}^{'\parallel}\right)
=
j\omega \mathbf{D'}^{\parallel c}_{m}  \left(\bar{\bar{\mathbf{\Gamma}}}\cdot{{\mathbf{r}}}^{'\parallel}\right) 
+
\left|m\right|\hat{\phi}\times\mathbf{H'}^{\parallel c}_{m}  \left(\bar{\bar{\mathbf{\Gamma}}}\cdot{{\mathbf{r}}}^{'\parallel}\right).
\label{eq:ap_NME_AL_mod_TE_pol}
\end{flalign}
Using the identity (\ref{eq:ap_iden_PML}), we can rewrite (\ref{eq:ap_NME_FL_mod_TE_pol})$-$(\ref{eq:ap_NME_AL_mod_TE_pol}) as
\begin{flalign}
{\nabla}'^{\parallel} \times
\left[
\bar{\bar{\mathbf{S}}}^{-1}\cdot\mathbf{E'}^{\parallel c}_{m}\left(\bar{\bar{\mathbf{\Gamma}}}\cdot{{\mathbf{r}}}^{'\parallel}\right)
\right]
&=
-j\omega
\left[
\left(\text{det}\bar{\bar{\mathbf{S}}}\right)^{-1}\bar{\bar{\mathbf{S}}}\cdot\mathbf{B'}^{\perp c}_{m}\left(\bar{\bar{\mathbf{\Gamma}}}\cdot{{\mathbf{r}}}^{'\parallel}\right)
\right],
\label{eq:ap_FL_maxwellian_TE_pol}
\\
{\nabla}'^{\parallel} \times
\left[
\bar{\bar{\mathbf{S}}}^{-1}\cdot\mathbf{E'}^{\perp c}_{m}\left(\bar{\bar{\mathbf{\Gamma}}}\cdot{{\mathbf{r}}}^{'\parallel}\right)
\right]
&=
-j\omega
\left[
\left(\text{det}\bar{\bar{\mathbf{S}}}\right)^{-1}\bar{\bar{\mathbf{S}}}\cdot\mathbf{B'}^{\parallel c}_{m}\left(\bar{\bar{\mathbf{\Gamma}}}\cdot{{\mathbf{r}}}^{'\parallel}\right)
\right]
\nonumber
\\
&\!\!\!\!\!\!\!\!\!\!-
\left|m\right|
\left[
\left(\text{det}\bar{\bar{\mathbf{S}}}\right)^{-1}\bar{\bar{\mathbf{S}}}\cdot\left\{\hat{\phi}\times\mathbf{E'}^{\parallel c}_{m}\left(\bar{\bar{\mathbf{\Gamma}}}\cdot{{\mathbf{r}}}^{'\parallel}\right)\right\}
\right],
\label{eq:ap_FL_maxwellian_TM_pol}
\\
{\nabla}'^{\parallel} \times
\left[
\bar{\bar{\mathbf{S}}}^{-1}\cdot\mathbf{H'}^{\parallel c}_{m}\left(\bar{\bar{\mathbf{\Gamma}}}\cdot{{\mathbf{r}}}^{'\parallel}\right)
\right]
&=
j\omega
\left[
\left(\text{det}\bar{\bar{\mathbf{S}}}\right)^{-1}\bar{\bar{\mathbf{S}}}\cdot\mathbf{D'}^{\perp c}_{m}\left(\bar{\bar{\mathbf{\Gamma}}}\cdot{{\mathbf{r}}}^{'\parallel}\right)
\right],
\label{eq:ap_AL_maxwellian_TM_pol}
\\
{\nabla}'^{\parallel} \times
\left[
\bar{\bar{\mathbf{S}}}^{-1}\cdot\mathbf{H'}^{\perp c}_{m}\left(\bar{\bar{\mathbf{\Gamma}}}\cdot{{\mathbf{r}}}^{'\parallel}\right)
\right]
&=
j\omega
\left[
\left(\text{det}\bar{\bar{\mathbf{S}}}\right)^{-1}\bar{\bar{\mathbf{S}}}\cdot\mathbf{D'}^{\parallel c}_{m}\left(\bar{\bar{\mathbf{\Gamma}}}\cdot{{\mathbf{r}}}^{'\parallel}\right)
\right]
\nonumber
\\
&\!\!\!\!\!\!\!\!\!\!+
\left|m\right|
\left[
\left(\text{det}\bar{\bar{\mathbf{S}}}\right)^{-1}\bar{\bar{\mathbf{S}}}\cdot\left\{\hat{\phi}\times\mathbf{H'}^{\parallel c}_{m}\left(\bar{\bar{\mathbf{\Gamma}}}\cdot{{\mathbf{r}}}^{'\parallel}\right)\right\}
\right].
\label{eq:ap_AL_maxwellian_TE_pol}
\end{flalign}   
We can further verify the identity below
\begin{flalign}
\left(\text{det}\bar{\bar{\mathbf{S}}}\right)^{-1}\bar{\bar{\mathbf{S}}}\cdot\left\{\hat{\phi}\times\mathbf{E'}^{\parallel c}_{m}\left(\bar{\bar{\mathbf{\Gamma}}}\cdot{{\mathbf{r}}}^{'\parallel}\right)\right\}
&=
\hat{\phi}\times\left[\bar{\bar{\mathbf{S}}}^{-1}\cdot\mathbf{E'}^{\parallel c}_{m}\left(\bar{\bar{\mathbf{\Gamma}}}\cdot{{\mathbf{r}}}^{'\parallel}\right)\right],
\label{eq:ap_ident_1_PML}
\\
\left(\text{det}\bar{\bar{\mathbf{S}}}\right)^{-1}\bar{\bar{\mathbf{S}}}\cdot\left\{\hat{\phi}\times\mathbf{H'}^{\parallel c}_{m}\left(\bar{\bar{\mathbf{\Gamma}}}\cdot{{\mathbf{r}}}^{'\parallel}\right)\right\}
&=
\hat{\phi}\times\left[\bar{\bar{\mathbf{S}}}^{-1}\cdot\mathbf{H'}^{\perp c}_{m}\left(\bar{\bar{\mathbf{\Gamma}}}\cdot{{\mathbf{r}}}^{'\parallel}\right)\right].
\label{eq:ap_ident_2_PML}
\end{flalign}
and introduce a new set of fields defined as
\begin{flalign}
\mathbf{E'}^{a}_{m}\left( {\mathbf{r}}^{'\parallel}\right)&=\bar{\bar{\mathbf{S}}}^{-1}\cdot\mathbf{E'}^{c}_{m}\left(\bar{\bar{\mathbf{\Gamma}}}\cdot{{\mathbf{r}}}^{'\parallel}\right),
\label{eq:ap_E_analytic_PML}
\\
\mathbf{H'}^{a}_{m}\left( {\mathbf{r}}^{'\parallel}\right)&=\bar{\bar{\mathbf{S}}}^{-1}\cdot\mathbf{H'}^{c}_{m}\left(\bar{\bar{\mathbf{\Gamma}}}\cdot{{\mathbf{r}}}^{'\parallel}\right),
\label{eq:ap_H_analytic_PML}
\\
\mathbf{D'}^{a}_{m}\left( {\mathbf{r}}^{'\parallel}\right)&=
\left(\text{det}\bar{\bar{\mathbf{S}}}\right)^{-1}
\bar{\bar{\mathbf{S}}}\cdot\mathbf{D'}^{c}_{m}\left(\bar{\bar{\mathbf{\Gamma}}}\cdot{{\mathbf{r}}}^{'\parallel}\right),
\label{eq:ap_D_analytic_PML}
\\
\mathbf{B'}^{a}_{m}\left( {\mathbf{r}}^{'\parallel}\right)&=
\left(\text{det}\bar{\bar{\mathbf{S}}}\right)^{-1}
\bar{\bar{\mathbf{S}}}\cdot\mathbf{B'}^{c}_{m}\left(\bar{\bar{\mathbf{\Gamma}}}\cdot{{\mathbf{r}}}^{'\parallel}\right),
\label{eq:ap_B_analytic_PML}
\end{flalign}
so that, by substituting (\ref{eq:ap_E_analytic_PML})$-$(\ref{eq:ap_B_analytic_PML}) back into (\ref{eq:ap_FL_maxwellian_TE_pol})$-$(\ref{eq:ap_AL_maxwellian_TE_pol}), and utilizing the identities (\ref{eq:ap_ident_1_PML}) and (\ref{eq:ap_ident_2_PML}), we finally obtain
\begin{flalign}
{\nabla}'^{\parallel} \times
\mathbf{E'}^{\parallel a}_{m}\left({\mathbf{r}}^{'\parallel}\right)
&=
-j\omega
\mathbf{B'}^{\perp a}_{m}\left({\mathbf{r}}^{'\parallel}\right),
\\
{\color{black}
{\nabla}'^{\parallel} \times
\mathbf{E'}^{\perp a}_{m}\left({\mathbf{r}}^{'\parallel}\right)
}
&
{\color{black}
=
-j\omega
\mathbf{B'}^{\parallel a}_{m}\left({\mathbf{r}}^{'\parallel}\right)
+
\left|m\right|
\mathbf{E'}^{\parallel a}_{m}\left({\mathbf{r}}^{'\parallel}\right)\times\hat{\phi},
}
\\
{\nabla}'^{\parallel} \times
\mathbf{H'}^{\parallel a}_{m}\left({\mathbf{r}}^{'\parallel}\right)
&=
j\omega
\mathbf{D'}^{\perp a}_{m}\left({\mathbf{r}}^{'\parallel}\right)
,
\\
{\color{black}
{\nabla}'^{\parallel} \times
\mathbf{H'}^{\perp a}_{m}\left({\mathbf{r}}^{'\parallel}\right)
}
&
{\color{black}
=
j\omega
\mathbf{D'}^{\parallel a}_{m}\left({\mathbf{r}}^{'\parallel}\right)
-
\left|m\right|
\mathbf{H'}^{\parallel a}_{m}\left({\mathbf{r}}^{'\parallel}\right)\times\hat{\phi}.
}
\end{flalign}
with
\begin{flalign}
\mathbf{D'}^{a}_{m}\left({{\mathbf{r}}}^{'\parallel}\right)
&=
\left[
\left(\text{det}\bar{\bar{\mathbf{S}}}\right)^{-1}
\left\{
\bar{\bar{\mathbf{S}}}\cdot
\bar{\bar{\epsilon}}'\left(\omega\right)\cdot
\bar{\bar{\mathbf{S}}}
\right\}
\right]
\cdot\mathbf{E'}^{a}_{m}\left({{\mathbf{r}}}^{'\parallel}\right),
\\
\mathbf{B'}^{a}_{m}\left({{\mathbf{r}}}^{'\parallel}\right)
&=
\left[
\left(\text{det}\bar{\bar{\mathbf{S}}}\right)^{-1}
\left\{
\bar{\bar{\mathbf{S}}}\cdot
\bar{\bar{\mu}}'\left(\omega\right)\cdot
\bar{\bar{\mathbf{S}}}
\right\}
\right]
\cdot\mathbf{H'}^{a}_{m}\left({{\mathbf{r}}}^{'\parallel}\right).
\end{flalign}
The above expressions show that $\mathbf{E'}^{a}_{m}$, $\mathbf{H'}^{a}_{m}$, $\mathbf{D'}^{a}_{m}$, and $\mathbf{B'}^{a}_{m}$ obey Maxwell's equations in an equivalent PML medium with constitutive parameters given by
\begin{flalign}
\bar{\bar{\epsilon}}^{\text{PML}}
&=
\left[\left(\text{det}\bar{\bar{\mathbf{S}}}\right)^{-1}
\left\{
\bar{\bar{\mathbf{S}}}\cdot
\bar{\bar{\epsilon}}'\left(\omega\right)\cdot
\bar{\bar{\mathbf{S}}}
\right\}
\right],
\label{eq:ap_eps_PML_final}
\\
\bar{\bar{\mu}}^{\text{PML}}
&=
\left[
\left(\text{det}\bar{\bar{\mathbf{S}}}\right)^{-1}
\left\{
\bar{\bar{\mathbf{S}}}\cdot
\bar{\bar{\mu}}'\left(\omega\right)\cdot
\bar{\bar{\mathbf{S}}}
\right\}
\right].
\label{eq:ap_mu_PML_final}
\end{flalign}

As an example, consider a background medium
with
\begin{flalign}
\bar{\bar{\epsilon}}\left(\omega\right)&=
\left[
\begin{matrix}
    \epsilon_{\rho}\left(\omega\right) & 0 & 0 \\
    0 & \epsilon_{\phi}\left(\omega\right) & 0 \\
    0 & 0 & \epsilon_{z}\left(\omega\right)
\end{matrix}
\right],
\\
\bar{\bar{\mu}}\left(\omega\right)&=
\left[
\begin{matrix}
    \mu_{\rho}\left(\omega\right) & 0 & 0 \\
    0 & \mu_{\phi}\left(\omega\right) & 0 \\
    0 & 0 & \mu_{z}\left(\omega\right)
\end{matrix}
\right],
\end{flalign}
with
$\epsilon_{\rho}\left(\omega\right)=\epsilon_{\phi}\left(\omega\right) = \epsilon_{z}\left(\omega\right) = \left(1+\frac{\sigma_{m}}{j\omega \epsilon_{0}}\right)$,
corresponding to a lossy, isotropic, homogeneous medium. After the TO-based mapping, we obtain
\begin{flalign}
\bar{\bar{\epsilon}}'\left(\omega\right)&=\bar{\bar{\epsilon}}\left(\omega\right)\cdot \bar{\bar{\mathbf{R}}}_{\epsilon}=
\left[
\begin{matrix}
    \epsilon_{\rho}\left(\omega\right)\rho & 0 & 0 \\
    0 & \frac{\epsilon_{\phi}\left(\omega\right)}{\rho} & 0 \\
    0 & 0 & \epsilon_{z}\left(\omega\right)\rho
\end{matrix}
\right],
\\
\bar{\bar{\mu}}'\left(\omega\right)&=\bar{\bar{\mu}}\left(\omega\right)\cdot \bar{\bar{\mathbf{R}}}_{\mu}=
\left[
\begin{matrix}
    \mu_{\rho}\left(\omega\right)\rho & 0 & 0 \\
    0 & \frac{\mu_{\phi}\left(\omega\right)}{\rho} & 0 \\
    0 & 0 & \mu_{z}\left(\omega\right)\rho
\end{matrix}
\right],
\end{flalign}
As a result, by using (\ref{eq:ap_eps_PML_final}) and (\ref{eq:ap_mu_PML_final}), the elements of the resulting PML constitutive tensor write as:
\begin{flalign}
&\epsilon_{\rho}^{\text{PML}}\left(\omega\right)=
\epsilon_{0}
\left(1+\frac{\sigma_{m}}{j\omega \epsilon_{0}}\right) 
\frac{\left( j\omega \epsilon_{0}+\sigma_{\rho}^{\text{PML}}\right)}
{\left( j\omega \epsilon_{0}+\sigma_{z}^{\text{PML}}\right)},
\\
&\epsilon_{\phi}^{\text{PML}}\left(\omega\right)=
\epsilon_{0}
\left(1+\frac{\sigma_{m}}{j\omega \epsilon_{0}}\right) 
\frac{ \left({ j\omega \epsilon_{0}}\right)^{2}}{
\left( j\omega \epsilon_{0}+\sigma_{\rho}^{\text{PML}}\right)
\left(  j\omega \epsilon_{0}+\sigma_{z}^{\text{PML}}\right)},
\\
&\epsilon_{z}^{\text{PML}}\left(\omega\right)=
\epsilon_{0}
\left(1+\frac{\sigma_{m}}{j\omega \epsilon_{0}}\right) 
\frac{\left( j\omega \epsilon_{0}+\sigma_{z}^{\text{PML}}\right)}
{\left( j\omega \epsilon_{0}+\sigma_{\rho}^{\text{PML}}\right)},
\\
&\mu_{\rho}^{\text{PML}}\left(\omega\right)=
\mu_{0}
\frac{\left( j\omega \epsilon_{0}+\sigma_{\rho}^{\text{PML}}\right)}
{\left( j\omega \epsilon_{0}+\sigma_{z}^{\text{PML}}\right)},
\\
&\mu_{\phi}^{\text{PML}}\left(\omega\right)=
\mu_{0}
\frac{ \left({ j\omega \epsilon_{0}}\right)^{2}}{
\left( j\omega \epsilon_{0}+\sigma_{\rho}^{\text{PML}}\right)
\left(  j\omega \epsilon_{0}+\sigma_{z}^{\text{PML}}\right)},
\\
&\mu_{z}^{\text{PML}}\left(\omega\right)=
\mu_{0}
\frac{\left( j\omega \epsilon_{0}+\sigma_{z}^{\text{PML}}\right)}
{\left( j\omega \epsilon_{0}+\sigma_{\rho}^{\text{PML}}\right)}.
\end{flalign}
where $\sigma_{\rho}^{\text{PML}}$ and $\sigma_{z}^{\text{PML}}$ are the artificial PML conductivities along $\rho$ and $z$ respectively.
The presence of $j\omega$ factors in the above Fourier-domain elements produce modifications in the corresponding field equations in the time-domain. These modifications are implemented using an auxiliary differential equation (ADE) approach as described in, e.g.,~\cite{donderici2008conformal,donderici2008mixed}.

{\color{black}
\section{Stability Conditions} \label{ap:stability_borfetd}
To determine the stability conditions, we express the field update in matrix form as
\begin{flalign}
\bar{\mathbf{w}}^{n+1}=\bar{\bar{\mathbf{G}}} \cdot\bar{\mathbf{w}}^{n}= \left( \bar{\bar{\mathbf{I}}} + \bar{\bar{\mathbf{T}}} \right)\cdot\bar{\mathbf{w}}^{n}
\end{flalign}
with
\begin{flalign}
\bar{\mathbf{w}}^{n}=
\left(
\begin{matrix}
     \left[\mathbb{B}_{m}^{\perp}\right]^{n-\frac{1}{2}} \\
     \left[\mathbb{B}_{m}^{\parallel}\right]^{n-\frac{1}{2}} \\
	\left[\mathbb{E}_{m}^{\perp}\right]^{n} \\
	\left[\mathbb{E}_{m}^{\parallel}\right]^{n} \\
\end{matrix}
\right),
~~~
\bar{\mathbf{w}}^{n+1}=
\left(
\begin{matrix}
     \left[\mathbb{B}_{m}^{\perp}\right]^{n+\frac{1}{2}} \\
     \left[\mathbb{B}_{m}^{\parallel}\right]^{n+\frac{1}{2}} \\
	\left[\mathbb{E}_{m}^{\perp}\right]^{n+1} \\
	\left[\mathbb{E}_{m}^{\parallel}\right]^{n+1} \\
\end{matrix}
\right),
\end{flalign}
and
\begin{flalign}
\bar{\bar{\mathbf{T}}}=
\left(
\begin{matrix}
\bar{\bar{\mathbf{0}}}_{N_{2}\times N_{2}}, & \bar{\bar{\mathbf{0}}}_{N_{2}\times N_{1}}, & \bar{\bar{\mathbf{0}}}_{N_{2}\times N_{0}}, & -\Delta t\left[\mathcal{D}_{\text{curl}}\right] \\
\bar{\bar{\mathbf{0}}}_{N_{1}\times N_{2}}, & \bar{\bar{\mathbf{0}}}_{N_{1}\times N_{1}}, & -\Delta t\left[\mathcal{D}_{\text{grad}}\right], & \Delta t \left|m\right|\bar{\bar{\mathbf{I}}}_{N_1\times N_1} \\
\bar{\bar{\mathbf{0}}}_{N_{0}\times N_{2}}, & \Delta t\bar{\bar{\mathbf{X}}}_{\text{TM}^{\phi}}, 
& 
\begin{matrix}
-{\Delta t}^{2}\bar{\bar{\mathbf{X}}}_{\text{TM}^{\phi}}\cdot\left[\mathcal{D}_{\text{grad}}\right]
\end{matrix}
, 
& {\Delta t}^{2} \left|m\right|\bar{\bar{\mathbf{X}}}_{\text{TM}^{\phi}} \\
\Delta t\bar{\bar{\mathbf{X}}}_{\text{TE}^{\phi}}, & -\Delta t\left|m\right|\bar{\bar{\mathbf{A}}}, & -{\Delta t}^{2}\left|m\right|\bar{\bar{\mathbf{A}}}\cdot\left[\mathcal{D}_{\text{grad}}\right], 
& 
\begin{matrix}
-{\Delta t}^{2}\bar{\bar{\mathbf{X}}}_{\text{TE}^{\phi}}\cdot\left[\mathcal{D}_{\text{curl}}\right]
-{\Delta t}^{2}\left|m\right|^{2}\bar{\bar{\mathbf{A}}}
\end{matrix}
\end{matrix}
\right),
\end{flalign}
where
\begin{flalign}
\bar{\bar{\mathbf{X}}}_{\text{TM}^{\phi}}&=\left(\left[\star_{\epsilon}\right]^{0\rightarrow0}\right)^{-1}\cdot\left[\mathcal{D}_{\text{grad}}\right]^{T}\cdot\left[\star_{\mu}^{-1}\right]^{1\rightarrow1}, \\
\bar{\bar{\mathbf{X}}}_{\text{TE}^{\phi}}&=\left(\left[\star_{\epsilon}\right]^{1\rightarrow1}\right)^{-1}\cdot\left[\mathcal{D}_{\text{curl}}\right]^{T}\cdot\left[\star_{\mu}^{-1}\right]^{2\rightarrow2}, \\
\bar{\bar{\mathbf{A}}}&=\left(\left[\star_{\epsilon}\right]^{1\rightarrow1}\right)^{-1}\cdot\left[\star_{\mu}^{-1}\right]^{1\rightarrow1}.
\end{flalign}
A necessary condition for stability is $\left|\lambda_{\bar{\bar{\mathbf{G}}}}\right|\leq1$ for all eigenvalues $\lambda_{\bar{\bar{\mathbf{G}}}}$ of $\bar{\bar{\mathbf{G}}}$~\cite{Wang_stability}.

When $m=0$, the field update equation becomes decoupled into two independent numerical integrators for $\text{TE}^{\phi}$ and $\text{TM}^{\phi}$ fields.
In this case, following~\cite{kim2011parallel}, we can easily obtain the stability criteria for both polarizations  in closed form as
\begin{flalign}
\Delta t_{\text{TE}^{\phi},m=0} \leq \frac{2}{\sqrt{\max\left(\lambda_{\mathbf{X_{\text{TE}^{\phi}}}\cdot\left[\mathcal{D}_{\text{curl}}\right]}\right)}},
\\
\Delta t_{\text{TM}^{\phi},m=0} \leq \frac{2}{\sqrt{\max\left(\lambda_{\mathbf{X_{\text{TM}^{\phi}}}\cdot\left[\mathcal{D}_{\text{grad}}\right]}\right)}},
\end{flalign}
where $\lambda_{\mathbf{X_{\text{TE}^{\phi}}}\cdot\left[\mathcal{D}_{\text{curl}}\right]}$ and $\lambda_{\mathbf{X_{\text{TM}^{\phi}}}\cdot\left[\mathcal{D}_{\text{grad}}\right]}$ denote the eigenvalues of $\mathbf{X_{\text{TE}^{\phi}}}\cdot\left[\mathcal{D}_{\text{curl}}\right]$ and $\mathbf{X_{\text{TM}^{\phi}}}\cdot\left[\mathcal{D}_{\text{grad}}\right]$ respectively. 

When $m\neq0$, we can simply represent $\bar{\bar{\mathbf{G}}}$ using $2 \times 2$ block matrices $\bar{\bar{\mathbf{X}}}$ and $\left[\mathcal{D}\right]$ as
\begin{flalign}
\bar{\bar{\mathbf{G}}}=\left[
\begin{matrix}
\bar{\bar{\mathbf{I}}}_{\left(N_2+N_1\right)\times\left(N_2+N_1\right)}, & -\Delta t\left[\mathcal{D}\right] \\
\Delta t\bar{\bar{\mathbf{X}}}, &  \bar{\bar{\mathbf{I}}}_{\left(N_0+N_1\right)\times\left(N_0+N_1\right)}-{\Delta t}^{2}\bar{\bar{\mathbf{X}}}\cdot\left[\mathcal{D}\right]\\
\end{matrix}
\right]
\end{flalign}
where
\begin{flalign}
\bar{\bar{\mathbf{X}}}=
\left[
\begin{matrix}
\bar{\bar{\mathbf{0}}}_{N_0\times N_2}, & \bar{\bar{\mathbf{X}}}_{\text{TM}^{\phi}} \\
\bar{\bar{\mathbf{X}}}_{\text{TE}^{\phi}}, & -\left|m\right|\bar{\bar{\mathbf{A}}} \\
\end{matrix}
\right],
\end{flalign}
and
\begin{flalign}
\left[\mathcal{D}\right]=
\left[
\begin{matrix}
\bar{\bar{\mathbf{0}}}_{N_2\times N_0}, & \left[\mathcal{D}_{\text{curl}}\right] \\
\left[\mathcal{D}_{\text{grad}}\right] & -\left|m\right|\bar{\bar{\mathbf{I}}}_{N_1\times N_1} \\
\end{matrix}
\right].
\end{flalign}
Therefore, the stability condition is similarly obtained as
\begin{flalign}
\Delta t_{m\neq0} \leq \frac{2}{\sqrt{\max\left(\lambda_{ \bar{\bar{\mathbf{X}}}\cdot\left[\mathcal{D}\right] }\right)}}
\end{flalign}
where $\lambda_{ \bar{\bar{\mathbf{X}}}\cdot\left[\mathcal{D}\right] }$ are the eigenvalues of $\bar{\bar{\mathbf{X}}}\cdot\left[\mathcal{D}\right]$. Note that in this case the maximum time step depends on the modal index magnitude $|m|$.
}
\section*{References}
\bibliography{FETD_BOR_bib}
\end{document}